\documentclass[aps,a4paper,twocolumn]{revtex4-2}

\usepackage{amsmath}
\usepackage{graphicx}
\usepackage{subfigure}
\usepackage[usenames,dvipsnames]{color}
\definecolor{darkblue}{RGB}{0,0,196}
\usepackage[colorlinks=true,linkcolor=darkblue,citecolor=darkblue,urlcolor=darkblue]{hyperref}
\usepackage{setspace}
\usepackage{multirow}
\usepackage{hyperref}
\usepackage{xcolor}

\hypersetup{
  colorlinks   = true, %Colours links instead of ugly boxes
  urlcolor     = blue, %Colour for external hyperlinks
  linkcolor    = blue, %Colour of internal links
  citecolor   = blue %Colour of citations
}
\usepackage{graphicx}
\usepackage{amsmath,bbm}
\usepackage{amssymb,bm}
\usepackage{yfonts}
\usepackage{comment}
\usepackage[normalem]{ulem}
\begin{document}

\title{Investigating radial flow-like effects via pseudorapidity and transverse spherocity dependence of particle production in pp collisions at the LHC}% Force line breaks with \\
\author{Aswathy Menon Kavumpadikkal Radhakrishnan$^1$}
% \email[]{aswathymenon2118ehep@gmail.com}
 \author{Suraj Prasad$^1$}
% \email[]{suraj.prasad@cern.ch}
 \author{Sushanta Tripathy$^2$}
 \author{Neelkamal Mallick$^1$}
% \email[]{neelkamal.mallick@cern.ch}
 \author{Raghunath Sahoo$^1$}\email[Corresponding author: ]{raghunath.sahoo@cern.ch}

\affiliation{$^1$Department of Physics, Indian Institute of Technology Indore, Simrol, Indore 453552, India}
\affiliation{$^2$CERN, Geneva 23 1211, Switzerland}

\date{\today}% It is always \today, today,

\begin{abstract}
Recent observations of quark-gluon plasma (QGP) like signatures in high multiplicity proton-proton (pp) collisions, have compelled the heavy-ion physics community to re-examine small collision systems for proper baseline studies. Event-shape-based studies in pp collisions have succeeded to a certain extent in identifying the rare events mimicking such heavy-ion-like behaviour. In this study, we incorporate PYTHIA8 and AMPT to study radial flow-like signatures in pp collisions at $\sqrt{s} = 13$ TeV as a function of transverse spherocity and pseudorapidity. The selection of softer events possibly carrying heavy-ion-like features is performed using the transverse spherocity event-shape observable. As the particle production mechanism in midrapidity differs greatly from the forward rapidity, a pseudorapidity-dependent study is meaningful. Keeping ALICE 3 upgrades at the LHC in mind, this study aims to demonstrate the transverse spherocity and pseudorapidity dependence of the mean transverse momentum, particle ratios, and kinetic freezeout parameters in pp collisions at $\sqrt{s}$ = 13 TeV using PYTHIA8. We observe that the isotropic events show enhanced radial-flow effects in all multiplicity classes, however, the jetty events show signatures of the radial flow-like effects only in high-multiplicity events. For the first time, we show the transverse spherocity and pseudorapidity dependence of partonic modification factor in pp collisions, which clearly shows that by choosing transverse spherocity, one can directly probe the radial flow-like effects in pp collisions at the LHC.

%The pseudorapidity dependence would help understand the scientific community for future upgrades. At the same time, the transverse spherocity will serve its purpose of identifying soft-QCD-dominated events in small collision systems. 

\end{abstract}
  
\maketitle

%\tableofcontents

\section{Introduction}
\label{intro}
One of the primary goals of heavy-ion collisions at ultra-relativistic energies is to probe the quantum chromodynamics (QCD) phase diagram. Such collisions at the Large Hadron Collider (LHC) and Relativistic Heavy-Ion Collider (RHIC) form a state of deconfined partons in thermal equilibrium called the quark-gluon plasma (QGP), which is believed to have existed a few microseconds after the Big Bang. It is nearly impossible to directly observe such a deconfined medium in heavy-ion collisions due to its short lifetime. However, there are a few indirect signatures that can signify the presence of such a medium. These QGP-signatures require a non-QGP baseline measurement to compare with. Traditionally, proton-proton (pp) collisions have been used as the baseline to study these signatures for last few decades; however, recent observations of ridge-like structures \cite{CMS:2015fgy,ALICE:2013snk}, observation of strangeness enhancement \cite{ALICE:2016fzo} and radial flow-like signatures~\cite{ALICE:2013wgn,ALICE:2016dei,CMS:2016fnw} in high multiplicity pp collisions have impelled the scientific community to re-examine the origin of such heavy-ion-like features in small collision systems.

In hydrodynamical models of heavy-ion collisions, the radial flow is believed to give a boost to the particles based on their transverse momentum ($p_{\rm T}$), where higher momentum particles get a greater boost compared to the lower momentum particles for a given particle species. This $p_{\rm T}$-dependent boost due to radial flow gives rise to the broadening of the $p_{\rm T}$--spectra of the particle, thus enhancing mean transverse momenta. The broadening also depends on the particle mass as particles with lower mass are less affected in the presence of radial flow as compared to the particles with higher mass \cite{Kisiel:2010xy}. This difference in broadening of transverse momentum spectra for particles of different masses leads to a peak-like structure around 2-3 GeV/$c$ in transverse momentum in the ratios of particle yields for different masses.  In Pb--Pb collisions, the central collisions are observed to have more such radial flow effect compared to the peripheral collisions \cite{Preghenella:2013qsv}. In contrast, kinetic theory-based models such as a multi-phase transport model (AMPT) are able to explain many of the experimental features of particle spectra and flow~\cite{Das:2022lqh, Altmann:2024icx, Mallick:2020ium, Mallick:2021wop, Mallick:2021hcs, Behera:2021zhi, Mallick:2022alr, Prasad:2022zbr, Behera:2023nwj, Mallick:2023vgi}. It includes radial and anisotropic flow, which provides crucial information on the collectivity of the system formed in heavy-ion collisions. Both elliptic and triangular flow from AMPT qualitatively explain the experimental results~\cite{Mallick:2020ium, Mallick:2021hcs, Mallick:2022alr, Prasad:2022zbr, Mallick:2023vgi}. In Refs.~\cite{Mallick:2021hcs, Mallick:2023vgi}, the authors study the number of constituent quark (NCQ) scaling of elliptic flow using AMPT, where a qualitative agreement between experiment and models can be observed. 

However, perturbative quantum chromodynamics (pQCD) inspired models such as PYTHIA~\cite{Sjostrand:2007gs} with the implementation of color reconnection (CR) and multi-partonic interactions (MPI) are seen to be capable of imitating the radial flow-like effects in pp collisions \cite{Ortiz:2016kpz, OrtizVelasquez:2013ofg, Ortiz:2022mfv}.
These radial flow-like effects in pp collisions using PYTHIA8 are shown to have a positive correlation with the number of multi-partonic interactions ($N_{\rm mpi}$). Events with a higher value of $N_{\rm mpi}$ can mimic enhanced radial flow-like effects~\cite{Ortiz:2020rwg} as compared to events with low $N_{\rm mpi}$. In experiments, it is impossible to measure $N_{\rm mpi}$ directly. However, event shape-based study using transverse spherocity is one such method to probe $N_{\rm mpi}$ as the correlation between transverse spherocity and $N_{\rm mpi}$ is found to be significant in simulations. Transverse spherocity, being an event-shape observable, is capable of separating the events based on their geometrical shapes~\cite{Cuautle:2014yda}. It can identify the soft-QCD-dominated isotropic events and the hard-QCD-dominated jetty events. This feature of transverse spherocity makes it a perfect tool for studying the emerging QGP-like signatures in small collision systems with special interest to the high $N_{\rm mpi}$ events~\cite{Cuautle:2014yda,Cuautle:2015kra,Salam:2009jx,Bencedi:2018ctm,Banfi:2010xy,Tripathy:2019blo,Khuntia:2018qox,Deb:2020ezw,Khatun:2019dml}.

 The particle production mechanism in high-energy nuclear and hadronic collisions can be affected by different effects, such as nuclear modification of partonic distribution function, multiple partonic scattering, possible parton saturation and radial flow, where these effects are expected to depend upon pseudorapidity and charged particle multiplicity of the produced hadrons~\cite{ALICE:2022uac, CMS:2016zzh, Kang:2012kc, ALICE:2013wgn, Bozek:2015swa}. A few studies on pseudorapidity dependence of particle production have been performed at the LHC~\cite{ALICE:2022uac, ALICE:2012mj, ALICE:2018vuu, CMS:2016zzh, CMS:2019isl}, and RHIC~\cite{BRAHMS:2016klg, PHOBOS:2003uzz, PHENIX:2018hho, PHENIX:2014fnc, PHENIX:2016cfs}. In Ref.~\cite{BRAHMS:2016klg}, the study of mean transverse radial flow velocity ($\langle \beta_{\rm T} \rangle$) and the kinetic freeze-out temperature ($T_{\rm kin}$) extracted from the simultaneous Boltzmann Gibbs blastwave function fit of the identified particles' $p_{\rm T}$--spectra in Cu--Cu collisions at $\sqrt{s_{\rm{NN}}} = 200$ GeV shows a rapidity dependence. Here, $\langle \beta_{\rm T} \rangle$ decreases towards the forward rapidity regions. This could be due to the availability of lower energy (and hence lower $\langle p_{\rm T} \rangle$) in the system at forward rapidity. In addition, using a multi-phase transport model in Pb--Pb collisions at LHC energies, both $\langle \beta_{\rm T} \rangle$ and $T_{\rm kin}$ are found to have a transverse spherocity dependence, indicating that the isotropic events can have more radial flow compared to the jetty events~\cite{Prasad:2021bdq, Tripathy:2019blo}. The studies mentioned above signify the importance of pseudorapidity and event-shape dependence of particle production in pp, p--Pb and Pb--Pb collisions. Thus, a systematic study of event-shape and pseudo-rapidity dependence of particle production in small collision systems can enable the scientific community to unveil the physical processes contributing to the observed heavy-ion-like phenomena in pp and p--Pb collisions~\cite{ALICE:2022uac}.

In this study, we employ transverse spherocity as an event shape observable to study the pseudorapidity dependence of radial flow-like effects in pp collisions at $\sqrt{s}=13$ TeV. We report observables sensitive to radial flow-like effects such as the mean transverse momentum, particle ratios, and kinetic freezeout parameters as a function of transverse spherocity and pseudorapidity. For this study, we use PYTHIA8, which explains the experimentally observed flow-like effects with good qualitative agreement and can provide a clear effect of MPI activity in different rapidity regions. A comparison with a multi-phase transport model is also shown for specific cases, which can help to make a model-independent agreement on the dependence of transverse spherocity and pseudorapidity on the reported observables. Another crucial point is that transverse spherocity is usually defined in the midrapidity ($|\eta|<$ 0.8) due to the current constraints of charged-particle measurement in the forward pseudorapidity at the LHC. However, with upcoming ALICE 3 upgrades, one can make measurements with particle identification for a wider pseudorapidity coverage, i.e., $|\eta|<4.0$~\cite{ALICE:2022wwr}. Thus, keeping ALICE 3 upgrades at the LHC in mind, this study aims to demonstrate the spherocity and pseudorapidity dependence of a few observables. Consequently, we define the transverse spherocity in a broader range of pseudorapidity \textit{i.e.} $S_{0}(|\eta|< 2.0)$.

The paper is organized as follows. We start with an adequate introduction in Sec.~\ref{intro}, then discuss the event generation using PYTHIA8 and AMPT as well as transverse spherocity in Sec.~\ref{pythiasphero}, and later discuss the observables, and results in Sec.~\ref{results}. Finally, we summarise our findings in Sec.~\ref{summary}.

\section{Event generation and methodology}
\label{pythiasphero}

In this section, we discuss the event generation model, specific tunes, and analysis methodology. We start the description with the pQCD-inspired PYTHIA8 model and kinetic-theory based AMPT model. We then discuss transverse spherocity, which is the event-shape observable.

\subsection{PYTHIA8}
PYTHIA is a widely used perturbative QCD-inspired Monte Carlo event generator to simulate hadronic, leptonic, and heavy-ion collisions with emphasis on physics related to small collision systems like pp collisions \cite{Sjostrand:2007gs}. PYTHIA8 involves soft and hard QCD processes and contains the models for initial and final state parton showers, multiple parton-parton interactions, beam remnants, string fragmentation, and particle decays. In our present study, we have implemented the default 4C tune of PYTHIA8 (version 8.308) with the MPI and CR mechanisms involving soft-QCD processes and hadronic decays \cite{Corke:2010yf,Andersson:1983ia}. The particles originating from the MPIs and the beam remnants form the underlying event. The Lund string fragmentation model performs the hadronisation of these partons \cite{Andersson:1983ia}. The CR picture makes sure that the string between the partons is arranged in such a way that the total string length is reduced, which in turn leads to reduced particle multiplicity of the event. The 4C tune \cite{Corke:2010yf} based on MPI and CR is able to reproduce many features of the experimental pp collisions data reasonably well~\cite{Ortiz:2016kpz, OrtizVelasquez:2013ofg}.

\subsection{A multi-phase transport model (AMPT)}

AMPT is a kinetic theory-based model and possesses four main components as follows~\cite{Zhang:1999mqa, Zhang:2000bd, Zhang:2000nc, Lin:2001cx, Lin:2001yd, Pal:2001zw, Lin:2001zk, Zhang:2002ug, Pal:2002aw, Lin:2002gc, Lin:2003ah, Lin:2003iq, Wang:1991hta, Zhang:1997ej, He:2017tla, Li:2001xh, Greco:2003mm}.

\begin{enumerate}
    \item Initialisation of collisions: The collisions in the AMPT model are initialized by HIJING where the production cross-sections of mini-jet partons and excited strings are calculated for pp collisions~\cite{Wang:1991hta}.

    \item Parton transport: In the default version of AMPT, the produced minijet partons are transported through Zhang's parton cascade model~\cite{Zhang:1997ej}. In the string melting version, excited strings also participate in Zhang's parton cascade.

    \item Hadronisation: In the default version of AMPT, the transported partons are hadronized using the Lund string fragmentation model. In the string melting version, the quark coalescence mechanism is used for the hadronization of the transported partons~\cite{Lin:2001zk, He:2017tla}.

    \item Hadron Transport: The produced hadrons are transported through meson-meson, meson-baryon, and baryon-baryon interactions using a relativistic transport model~\cite{Li:2001xh, Greco:2003mm}.

\end{enumerate}

The string-melting mode of AMPT is able to reproduce many of the experimental features of particle flow and $p_{\rm T}$-spectra~\cite{Fries:2003vb, Fries:2003kq}. In this study, we use the string melting mode of AMPT (version 2.26t9b). We use the Lund symmetric splitting function parameters as $a=0.5$ and $b=0.9$ to match the experimental $p_{\rm T}$-spectra. In addition, the parton cross-section is set to 3 mb with a parton screening mass of 2.265 fm$^{-1}$ and alpha in the parton cascade as 0.33. All other settings of AMPT are kept the same as in Ref.~\cite{Tripathy:2018bib}. With these settings in AMPT, we simulate pp collisions at $\sqrt{s}=13$ TeV for this study.

Figure~\ref{fig:datacomp} (in the appendix) shows the comparison of the above-mentioned tunes of PYTHIA8 with the AMPT and ALICE measurements from Ref.~\cite{ALICE:2015qqj} for $p_{\rm T}$-spectra of minimum bias pp collisions at $\sqrt{s}$ = 13 TeV. As shown, the PYTHIA8-based 4C tune overestimates the experimental data, where it keeps a good qualitative agreement with the experimental measurements within 10\% uncertainties. On the other hand, the $p_{\rm T}$-spectra from AMPT underestimate the experimental curve, where it gives a good description only in the intermediate to high $p_{\rm T}$ region.

\subsection{Transverse Spherocity}
Transverse spherocity is a well-established event shape classifier in pp collisions, capable of separating the pQCD dominated hard events from the soft-QCD events. It is also observed that the selections based on the transverse spherocity, indeed, can identify the events with enhanced production of strange hadrons~\cite{Nassirpour:2022ihm}, similar to radial flow effects \cite{Cuautle:2015kra} in pp collisions.

%Due to its unique feature to identify and separate the rare pp collisions, transverse spherocity-based selections are useful for studying these QGP signatures.

Transverse spherocity, an infrared and co-linear safe quantity, is defined for a unit vector $\hat{n} (n_{\rm T},0)$ in the transverse plane as~\cite{Cuautle:2014yda},

\begin{equation}
    S_{0}=\frac{\pi^2}{4}\min_{\hat{n}}\Bigg(\frac{\sum_{i=1}^{N_{\rm had}}|\bf{p_{\rm T}} \times \hat{n}|}{\sum_{i=1}^{N_{\rm had}}|\bf{p_{\rm T}}|}\Bigg)^{2}
    \label{eq:spherodefn}
\end{equation}

Here, $\hat{n}$ is chosen such that the ratio within parenthesis is minimized for a given event. The summations run over all the charged hadrons in an event, and $N_{\rm had}$ is the total number of charged hadrons. Multiplication of $\pi^{2}/4$ in Eq.~\ref{eq:spherodefn} makes $S_0$ to lie between 0 and 1. The extreme limits of $S_0$ represent specific configurations of the produced particles on the transverse plane. The lower limit of transverse spherocity ($S_0 \to 0$) characterizes hard jetty events with back-to-back (pencil-like) emission of particles, while the higher limit ($S_0 \to 1$) corresponds to soft isotropic events. In this study, transverse spherocity is determined using the charged hadrons in the pseudorapidity range, $|\eta| < 2.0$ having transverse momentum, $p_{\rm T} > 0.15$ GeV/$c$. We restrict our estimations of transverse spherocity only to the events having more than 5 charged hadrons within the mentioned kinematic region. The jetty and isotropic event classes are categorised by choosing transverse spherocity cuts corresponding to the lowest 20\% and highest 20\% events in the spherocity distribution, respectively. For the sake of simplicity, we may conveniently replace transverse spherocity as spherocity throughout the remaining text.

\begin{figure}[ht!]
\includegraphics[scale=0.42]{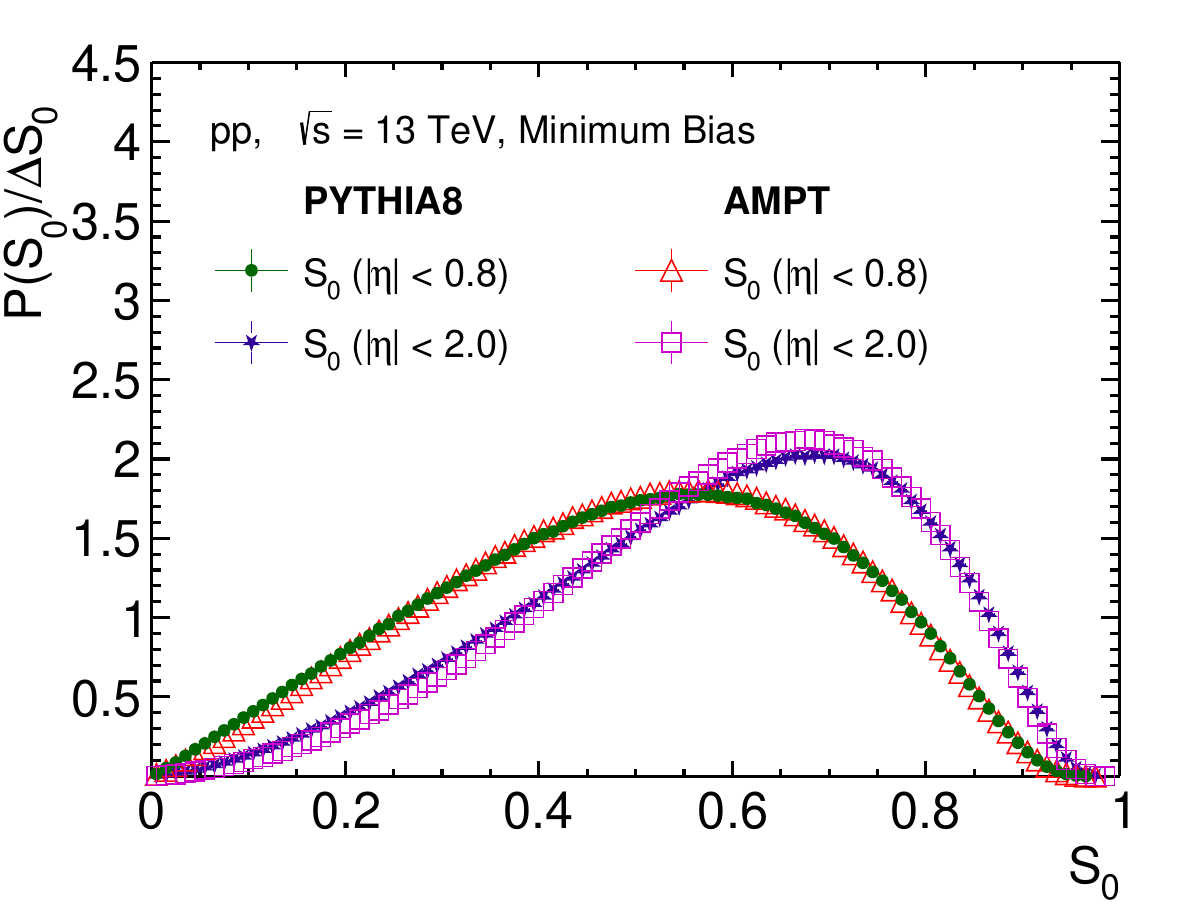}
\caption{(Color online) Transverse spherocity distribution ($S_{0}(|\eta|<0.8)$ and $S_{0}(|\eta|<2.0)$)  in pp collisions at $\sqrt{s}$ = 13 TeV using PYTHIA8 and AMPT.}
\label{fig:spherodist}
\end{figure}

\begin{figure}[ht!]
\includegraphics[scale=0.42]{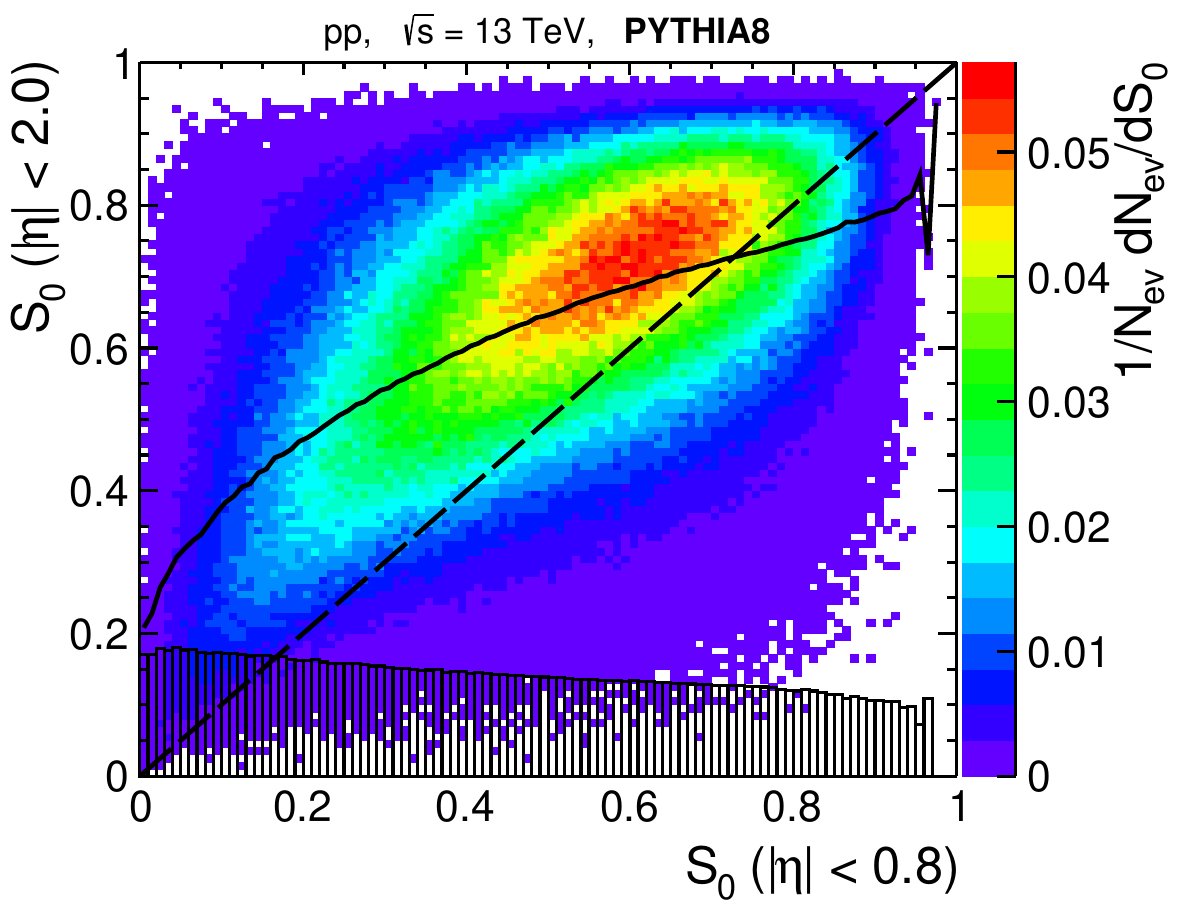}
\caption{(Color online) Correlation between $S_{0}$ measured in $|\eta|<0.8$ and $|\eta|<2.0$ in pp collisions at $\sqrt{s}=13$ TeV using PYTHIA8. The solid line represents the average value of the $y$-axis at a particular value of $x$-axis and the dashed line guides $S_{0}(|\eta|<0.8)=S_{0}(|\eta|<2.0)$. The bar representation in the figure shows the standard deviation value of the $y$-axis for each bin in $x$-axis.}
\label{fig:D02da}
\end{figure}

% \begin{figure}[ht!]
% \includegraphics[scale=0.42]{nMPI_0.8_S0_2.0.pdf}
% \caption{(Color online) Mean number of multipartonic interactions ($\langle N_{\rm mpi} \rangle$) as a function of transverse spherocity in pp collisions at $\sqrt{s}=13$ TeV using PYTHIA8.}
% \label{fig:nMPIS0}
% \end{figure}

\begin{figure}[ht!]
\includegraphics[scale=0.4]{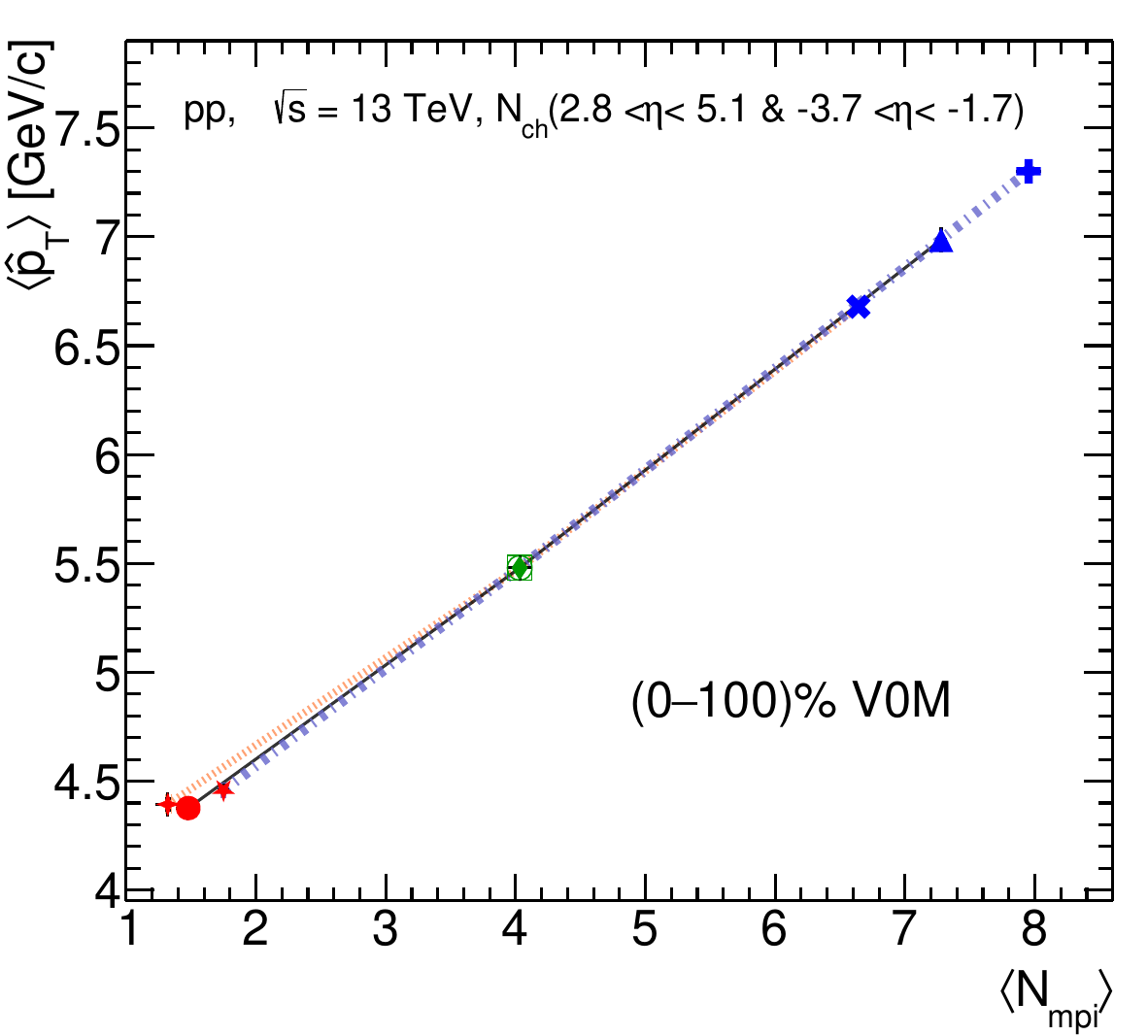}
\includegraphics[scale=0.4]{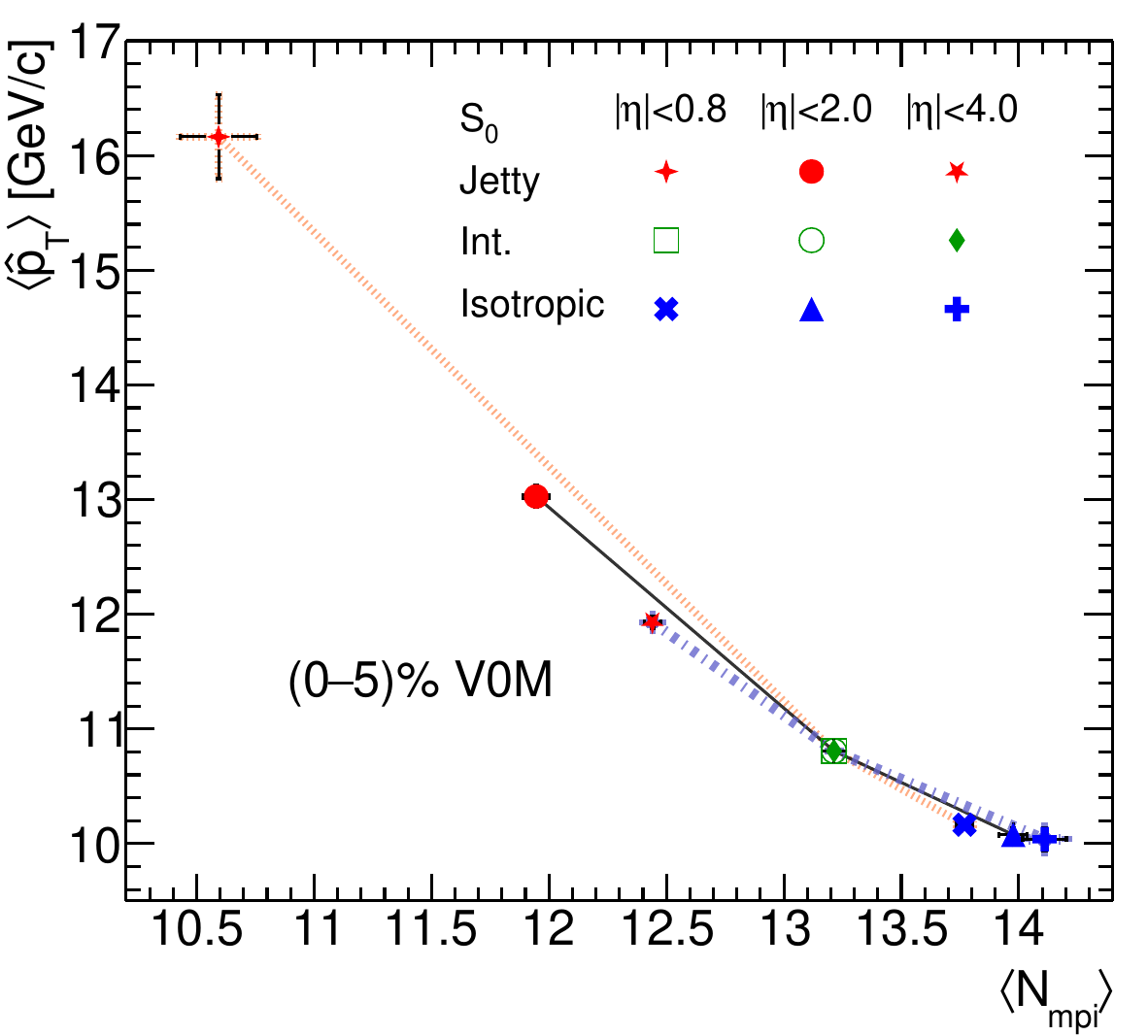}
\caption{(Color online) Correlation between $\langle\hat{p}_{\rm T}\rangle$ and $\langle N_{\rm mpi}\rangle$ as a function of $S_{0}(|\eta|<0.8)$, $S_{0}(|\eta|<2.0)$ and $S_{0}(|\eta|<4.0)$ in (0-100)\% V0M (upper) and (0-5)\% V0M (lower) in pp collisions at $\sqrt{s}=13$ TeV using PYTHIA8.}
\label{fig:3S0comppThatNmpi}
\end{figure}

\begin{figure}[ht!]
\includegraphics[scale=0.4]{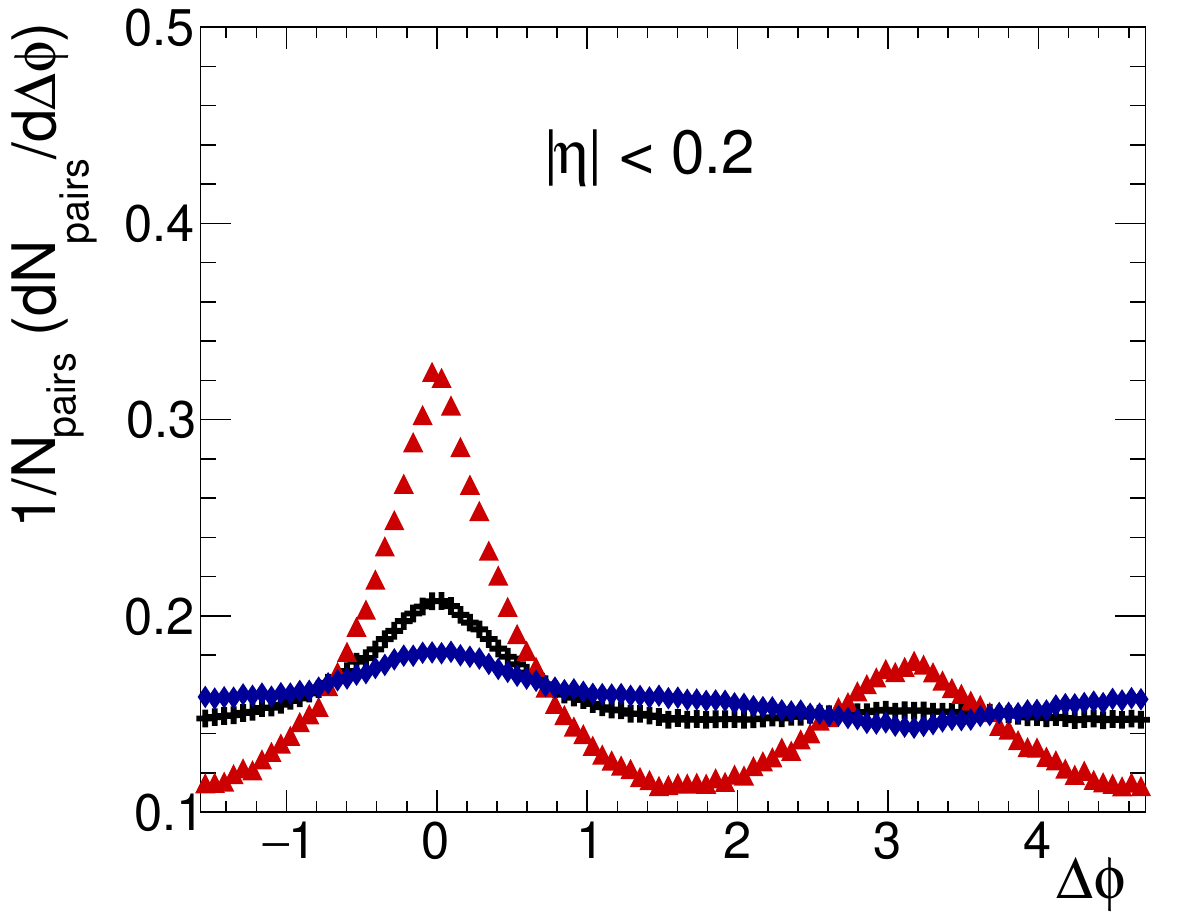}
\includegraphics[scale=0.4]{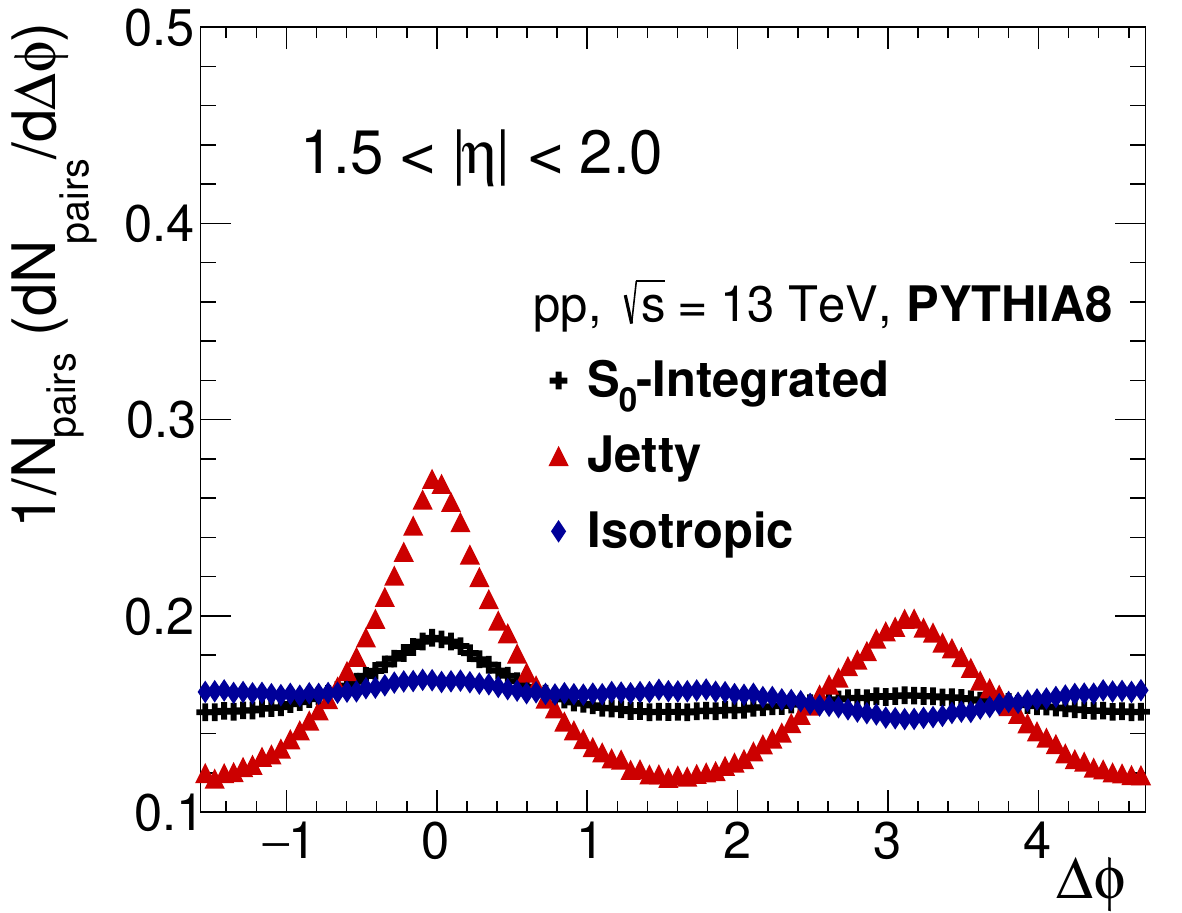}
\caption{(Color online) Spherocity dependence of $\Delta\phi$ distributions for charged-particles in $|\eta|<0.2$ (top) and $1.5<|\eta|<2.0$ (bottom) for $(0-100)\%$ V0M class in pp collisions at $\sqrt{s}=13$ TeV using PYTHIA8.}
\label{fig:DeltaphiNoleadcut}
\end{figure}

Figure~\ref{fig:spherodist} compares the transverse spherocity distributions for minimum bias pp collisions at $\sqrt{s}$ = 13 TeV using PYTHIA8 as well as AMPT with two kinematics cuts, one with spherocity defined for the particle tracks in the conventional pseudorapidity range, $|\eta| < 0.8$ and the other one with the spherocity in the pseudorapidity range $|\eta| < 2.0$. Here, the peak of $S_{0}$ distribution is shown to be shifted to a higher spherocity limit as we widen the rapidity region for transverse spherocity from $|\eta| < 0.8$ to $|\eta| < 2.0$ for both PYTHIA8 and AMPT simulations. The shift of the $S_0 (|\eta|<2.0)$ distribution to the right as compared to the $S_0 (|\eta|<0.8)$ curve is attributed to the inclusion of softer particles (low $p_{\rm T}$) as we move to a wider pseudorapidity range. In addition, it is also interesting to note that the transverse spherocity distribution for both $|\eta| < 0.8$ and $|\eta| < 2.0$ from PYTHIA8 matches the distributions from AMPT quite well.

Figure~\ref{fig:D02da} shows the correlation between two definitions of spherocity, \textit{i.e.,} between $S_{0}(|\eta|<0.8)$ and $S_{0}(|\eta|<2.0)$ from PYTHIA8; the solid line represents the average value of $y$-axis at a particular value of $x$-axis. The figure pictorially represents the shift in the spherocity values after the inclusion of a wider pseudorapidity region, \textit{i.e.}, from $|\eta| < 0.8$ to $|\eta| < 2.0$, for the estimation of spherocity. As one notices, the region above the $x=y$ line for $S_{0}(|\eta|<0.8)<0.7$ is largely populated. This indicates that majority of events having $S_{0}(|\eta|<0.8)<0.7$ acquire a larger $S_{0}(|\eta|<2.0)$ value. On the other hand, the events having $S_{0}(|\eta|<0.8)>0.7$ largely populate the region below $x=y$ line; however, the shift of mean and variance of $S_{0}(|\eta|<2.0)$ is small compared to the events having $S_{0}(|\eta|<0.8)<0.7$. This fairly indicates that the isotropic events are less affected by the change of the pseudorapidity window in the definition of transverse spherocity as compared to the jetty events.

\begin{table}[ht!]
\begin{tabular}{|c||c||c||c|c|}
\hline
\multirow{2}{*}{\bf{\shortstack{V0M \\ Percentile}}} & \multirow{2}{*}{\bf{$\langle dN_{\rm ch}/d\eta \rangle|_{|\eta|<2.0}$}} & \multicolumn{2}{c|}{\bf{$S_{0}$($|\eta|<2.0$) range}} \\\cline{3-4}
                        & & \bf{Jetty}  & \bf{Isotropic}  \\ \hline\hline
0 -- 1   &25.323 $\pm$ 0.017      & 0 - 0.675   & 0.865 - 1       \\ \hline
1 -- 5  &20.074 $\pm$ 0.007       & 0 - 0.645   & 0.845 - 1       \\ \hline
5 -- 10 &16.427 $\pm$ 0.006        & 0 - 0.615   & 0.835 - 1       \\ \hline
10 -- 20 &12.903 $\pm$ 0.004        & 0 - 0.585   & 0.815 - 1       \\ \hline
20 -- 30  &9.550 $\pm$ 0.003      & 0 - 0.535   & 0.795 - 1       \\ \hline
30 -- 40  &7.054 $\pm$ 0.003       & 0 - 0.485   & 0.765 - 1       \\ \hline
40 -- 50 &5.091 $\pm$ 0.002        & 0 - 0.425   & 0.735 - 1       \\ \hline
50 -- 70  &3.377 $\pm$ 0.001       & 0 - 0.355   & 0.695 - 1       \\ \hline
0 -- 100 &  6.702$\pm$ 0.002 & 0 - 0.425  & 0.765 - 1 \\ \hline

\end{tabular}
\caption{\label{spherocutPYTHIA} Transverse spherocity cuts for the jetty and isotropic events for different multiplicity classes in pp collisions at $\sqrt{s}$ = 13 TeV using PYTHIA8.} 
\end{table}

\begin{table}[ht!]
\begin{tabular}{|c||c||c||c|c|}
\hline
\multirow{2}{*}{\bf{\shortstack{V0M \\ Percentile}}} & \multirow{2}{*}{\bf{$\langle dN_{\rm ch}/d\eta \rangle|_{|\eta|<2.0}$}} & \multicolumn{2}{c|}{\bf{$S_{0}$($|\eta|<2.0$) range}} \\\cline{3-4}
                        & & \bf{Jetty}  & \bf{Isotropic}  \\ \hline\hline
0 -- 1   & 25.780 $\pm$ 0.018       & 0 - 0.662   & 0.858 - 1       \\ \hline
1 -- 5  & 19.792 $\pm$ 0.008        & 0 - 0.632   & 0.844 - 1       \\ \hline
5 -- 10 &  15.480 $\pm$ 0.006       & 0 -  0.600   & 0.828 - 1       \\ \hline
10 -- 20 & 11.472 $\pm$ 0.004       & 0 -  0.556   & 0.808 - 1       \\ \hline
20 -- 30  & 8.310 $\pm$ 0.003      & 0 - 0.512   & 0.786 - 1       \\ \hline
30 -- 40  &  6.368 $\pm$ 0.002     & 0 - 0.478   & 0.766 - 1       \\ \hline
40 -- 50 &  4.907 $\pm$ 0.002      & 0 - 0.446   & 0.748 - 1       \\ \hline
50 -- 70  &  3.600 $\pm$ 0.001     & 0 - 0.402   & 0.722 - 1       \\ \hline
0 -- 100 & 6.476 $\pm$ 0.002  & 0 - 0.435  & 0.765 - 1 \\ \hline

\end{tabular}
\caption{\label{spherocutAMPT} Transverse spherocity cuts for the jetty and isotropic events for different multiplicity classes in pp collisions at $\sqrt{s}$ = 13 TeV using AMPT.}
\end{table}

% In addition, the inclusion of the broader pseudorapidity bin for the definition of $S_0$ also affects its correlation with the $N_{\rm mpi}$, as shown in Fig.~\ref{fig:nMPIS0}. Here, $\langle N_{\rm mpi}\rangle$ is shown as a function of $S_0$ defined in two different pseudorapidity bins using PYTHIA8. Both the definitions of transverse spherocity are observed to possess a similar degree of correlation with $N_{\rm mpi}$. It is found that the spherocity defined in a narrower  $\eta$-cut achieves a higher $\langle N_{\rm mpi}\rangle$ value for the most isotropic events as compared to the spherocity defined in a broader $\eta$-cut. In contrast, the spherocity in the wider region can probe towards a lower value of $\langle N_{\rm mpi}\rangle$. This is expected as the transverse spherocity is correlated with the number of charged particles irrespective of the defined pseudorapidity region. Thus, the inclusion of a wider pseudorapidity region makes a semi-soft event become softer, while the hard events are not much affected by it. 
% This makes the event selection based on $S_{0} (|\eta|<2.0)$ a little insensitive to  higher values of $\langle N_{\rm mpi}\rangle$, while $S_{0} (|\eta|<0.8)$ sensitive to high values of $\langle N_{\rm mpi}\rangle$. In other words, $S_{0} (|\eta|<0.8)$ is a good probe to separate the isotropic events while $S_{0} (|\eta|<2.0)$ can handle the hard events better. 

Figure~\ref{fig:3S0comppThatNmpi} shows the correlation between average transverse momentum transfer of the hardest parton–parton interaction ($\langle\hat{p}_{\rm T}\rangle$) and average number of multi-partonic interactions ($\langle N_{\rm mpi}\rangle$) as a function of $S_{0}(|\eta|<0.8)$, $S_{0}(|\eta|<2.0)$ and $S_{0}(|\eta|<4.0)$ in (0-100)\% (upper) and (0-5)\% (lower) V0M classes in pp collisions at $\sqrt{s}=13$ TeV using PYTHIA8. For (0-100)\% V0M case, one finds a linear correlation between $\langle N_{\rm mpi}\rangle$ and $\langle\hat{p}_{\rm T}\rangle$. Here, $\langle\hat{p}_{\rm T}\rangle$ increases as one goes to isotropic events with increasing $\langle N_{\rm mpi}\rangle$. However, for (0-5)\% V0M class, which is dominated by both soft processes and multi-jet topologies leading to higher multiplicity, one finds an anti-correlation between $\langle N_{\rm mpi}\rangle$ and $\langle\hat{p}_{\rm T}\rangle$ as a function of transverse spherocity selection~\cite{ALICE:2022qxg, ALICE:2023plt}. $S_{0}(|\eta|<4.0)$ is found to isolate events with large $\langle N_{\rm mpi}\rangle$ in both (0-100)\% and (0-5)\% V0M classes. On the other hand, $S_{0}(|\eta|<0.8)$ is limited with a smaller value of $\langle N_{\rm mpi}\rangle$ in the isotropic events as compared to the other two definitions of transverse spherocity. In contrast, the advantage of using $S_{0}(|\eta|<2.0)$ as an event shape classifier is that it can probe events with higher $\langle N_{\rm mpi}\rangle$ as compared to $S_{0}(|\eta|<0.8)$, and is better at isolating hard events as compared to $S_{0}(|\eta|<4.0)$.

In this study, the charged particle multiplicity selection is made in the V0 detector acceptance region (V0M) of the ALICE experiment at the LHC, \textit{i.e.}, $-3.7 < \eta < -1.7$ and $2.8 < \eta < 5.1$ \cite{ALICE:2014sbx}. The  percentile cuts for the corresponding V0M selection from PYTHIA8 and AMPT are listed in the first column of Tab.~\ref{spherocutPYTHIA} and Tab.~\ref{spherocutAMPT} respectively. In addition, the table includes corresponding mean charged particle density measured in the mid-rapidity region, \textit{i.e.}, $|\eta| < 2.0$ along with the spherocity cuts for jetty and isotropic events in each V0M class. Here, the differential study for pseudorapidity is performed in six pseudorapidity regions, \textit{viz.}, $|\eta|<0.2$, $0.2\leq|\eta|<0.4$, $0.4\leq |\eta|<0.7$, $0.7\leq |\eta|<1$, $1\leq|\eta|<1.5$ and $1.5\leq|\eta|<2$. From Figs.~\ref{fig:spherodist},~\ref{fig:D02da}, and~\ref{fig:3S0comppThatNmpi}, it is evident that the spherocity estimated with a wider pseudorapidity cut behaves qualitatively similar to the midrapidity definition, and therefore, spherocity estimated with $|\eta|<2.0$ can be taken as the standard definition for the rest of results mentioned in this study. Exceptions, if any, shall be mentioned in the respective cases. The choice to widen the $\eta$ range for the estimation of $S_0$ for this particular study is only because this work aims to study the pseudorapidity dependence of radial flow-like effects. However, we do not compare the results obtained from the event-shape selection with the two definitions of transverse spherocity.

%Here onward, for all the event selection cuts, we use the transverse spherocity defined in $|\eta|<2.0$, \textit{i.e.,} $S_{0} (|\eta|<2.0)$ unless mentioned otherwise.

Finally, the event selection capability of spherocity, $S_0(|\eta|<2.0)$, for the jetty and isotropic events, can be evaluated by studying the relative azimuthal angle ($\Delta\phi$) distribution of all the charged particles with respect to the highest $p_{\rm T}$ charged hadron known as the trigger, which is calculated as $\Delta\phi = \phi_{\rm trig.}-\phi_{\rm track}$. Figure~\ref{fig:DeltaphiNoleadcut} shows the $\Delta\phi$ distributions for the charged-particles in two different pseudorapidity cuts, \textit{i.e.}, in $|\eta|<0.2$ and $1.5<|\eta|<2.0$, by considering the triggers in the respective $\eta$-cuts in $(0-100)\%$ V0M class from PYTHIA8. The jetty and isotropic event selections are performed using $S_0(|\eta|<2.0)$ for both plots. The plots are normalized by dividing the number of particle pairs ($N_{\rm pairs}$) in the respective $\eta$ ranges.
In both the pseudorapidity cases, the jetty events show larger near and away side peaks and the isotropic events show an almost flat (symmetric) distribution of particles in $\Delta\phi$ as compared to the $S_0$-integrated events. The presence of clear jet peaks in jetty events and a flat distribution in isotropic events in both the $\eta$-cuts, signify that the capability of spherocity to disentangle the jetty and isotropic events remains intact, and hence, it is a suitable candidate for the event selection as well as the study of pseudorapidity dependent properties of various observables which are described in the next section.

\section{Results and discussions}
\label{results}

In this section, we present the results of some of the observables, such as particle ratios, partonic modification factor, mean transverse momentum, kinetic freezeout properties, and transverse momentum crossing points, which are believed to be sensitive to the radial flow effects.

\subsection{Particle ratios}

\begin{figure*}[ht!]
\includegraphics[scale=0.42]{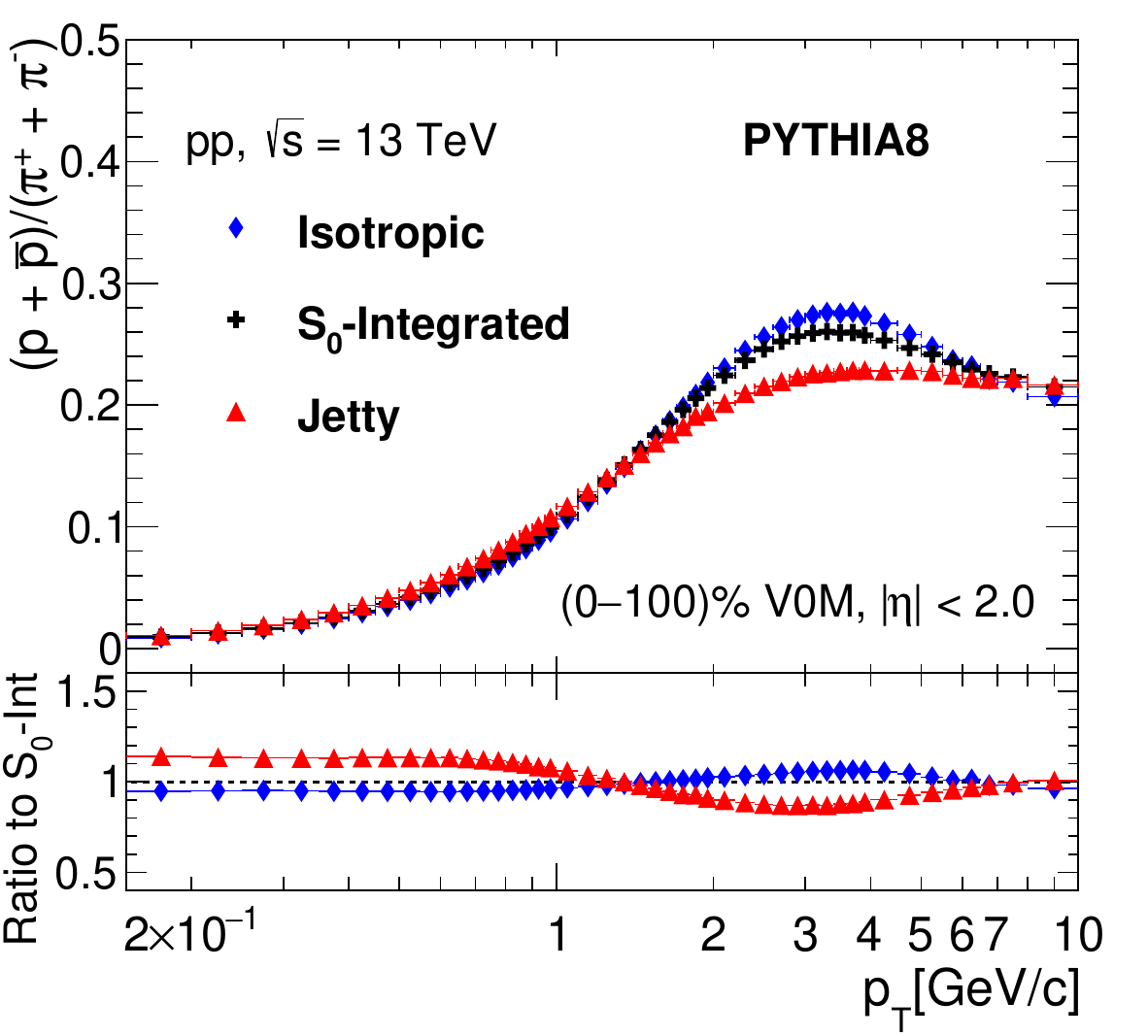}
\includegraphics[scale=0.42]{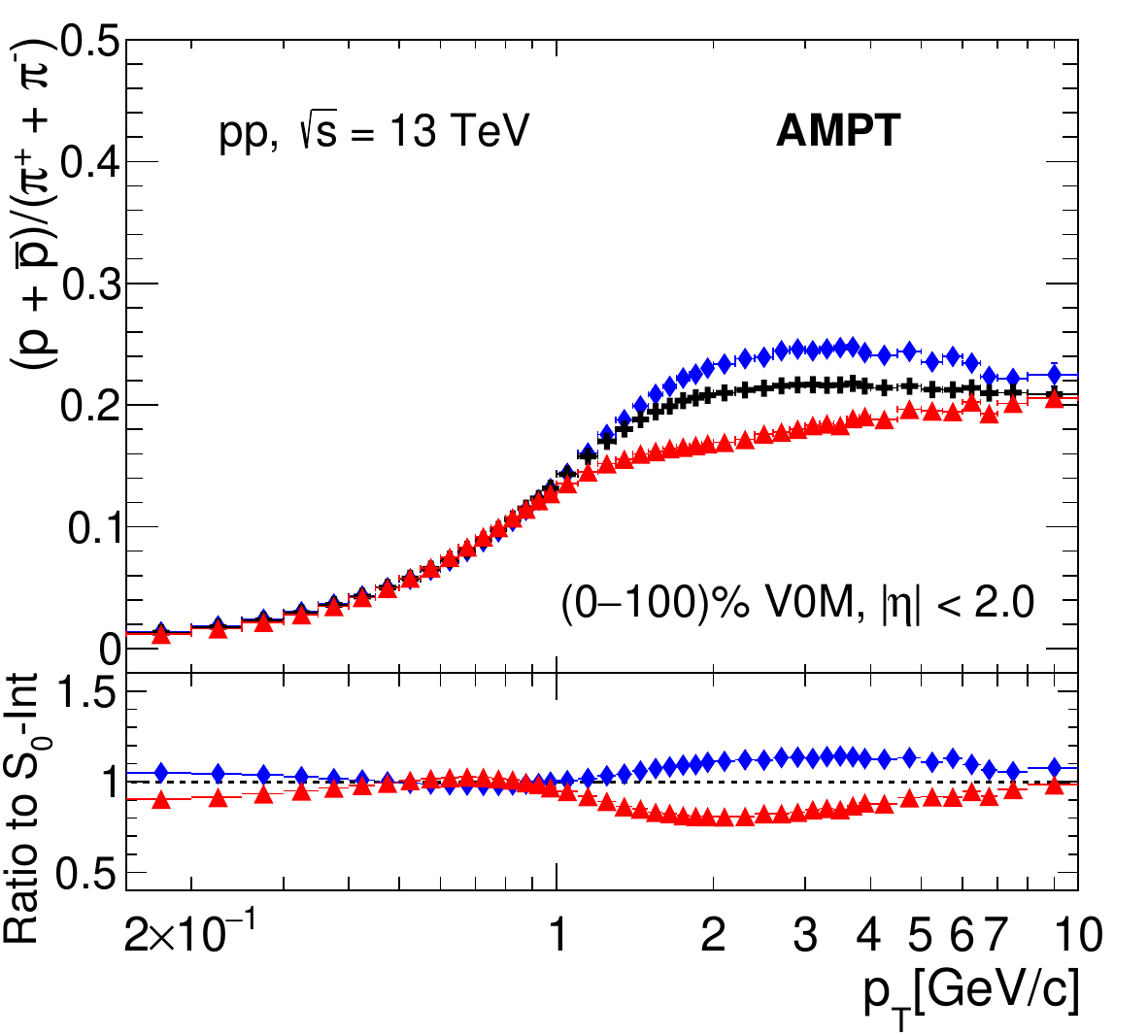}
\includegraphics[scale=0.42]{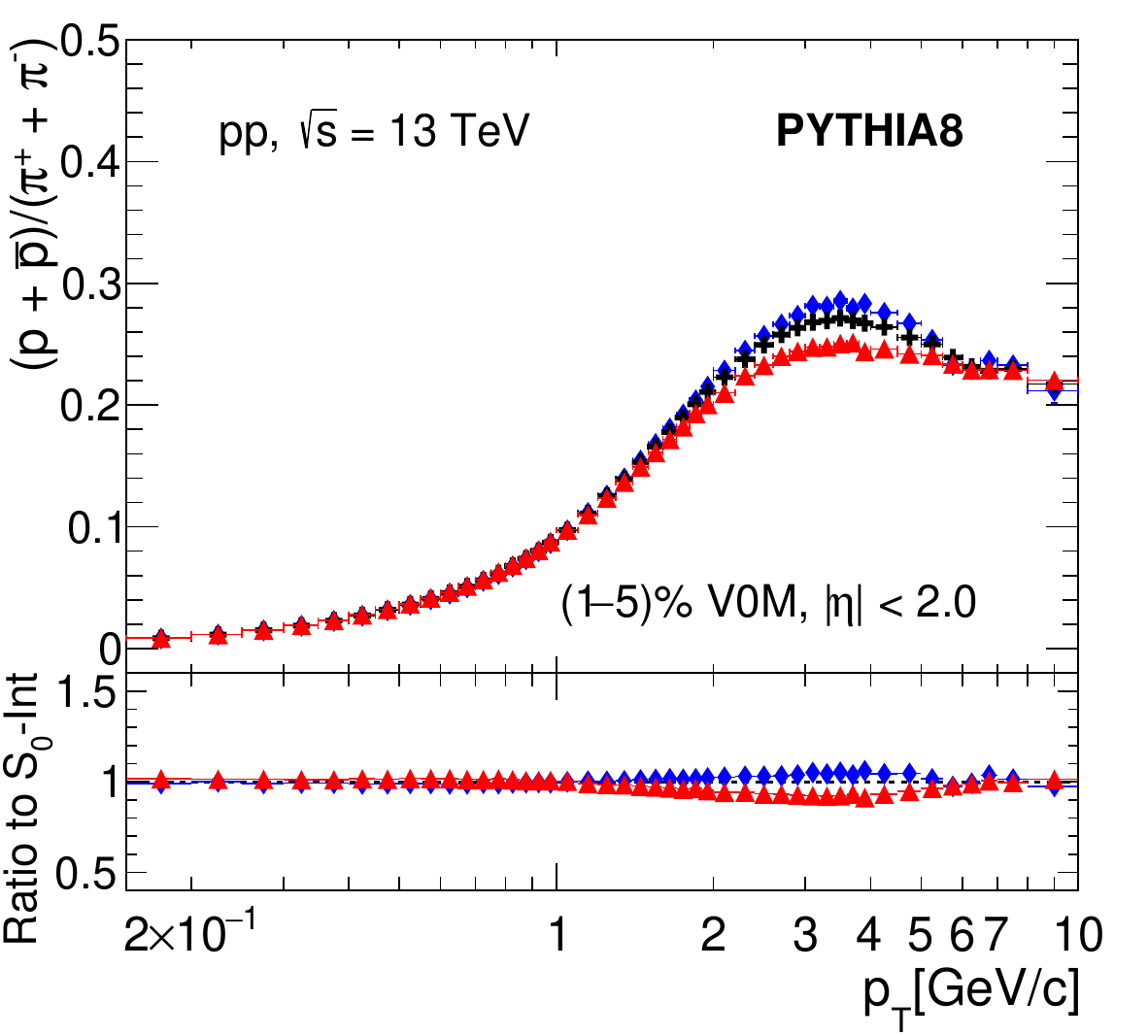}
\includegraphics[scale=0.42]{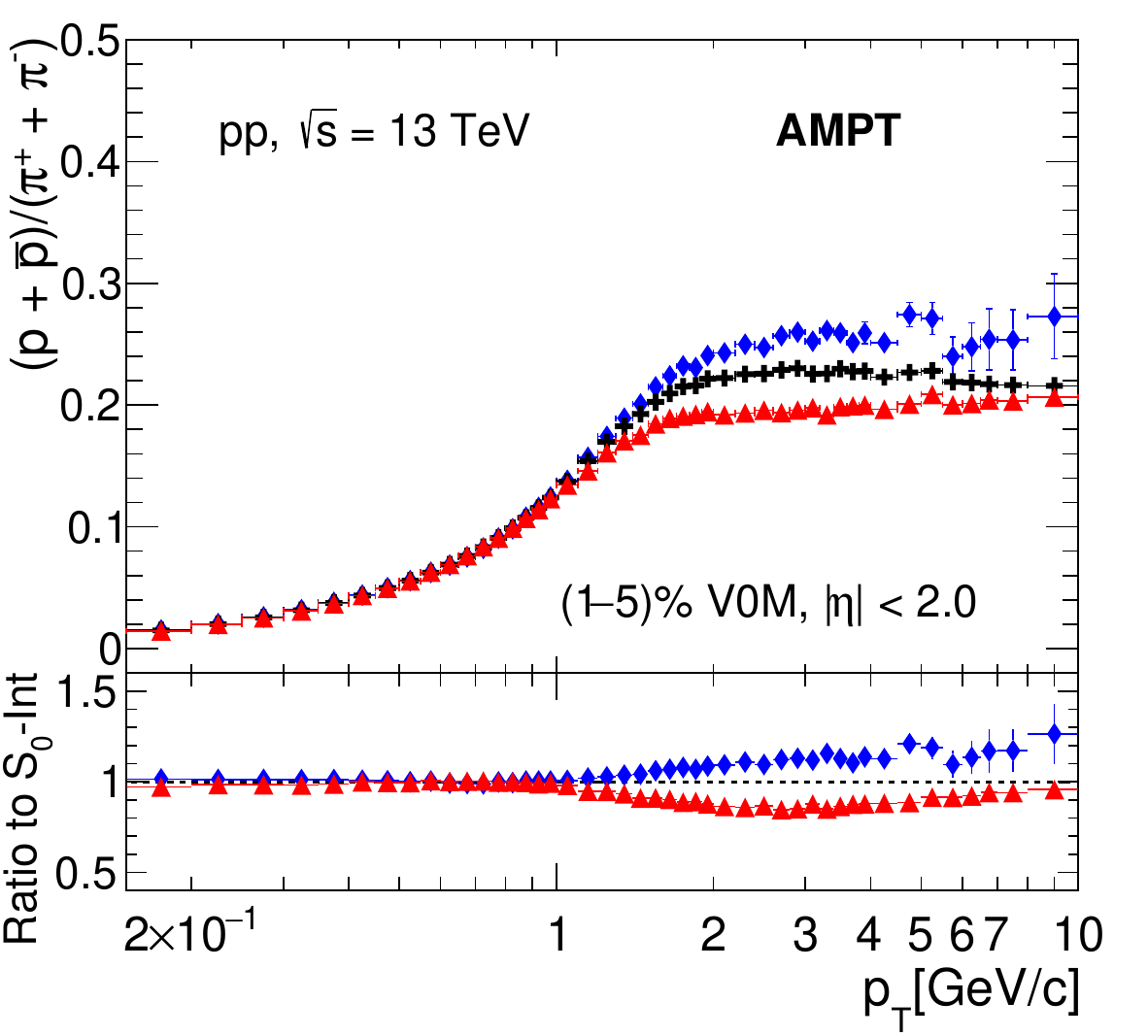}
\caption{(Color online) Proton-to-pion ratio as a function of transverse momentum at $|\eta|<2.0$ for different spherocity bins for $(0-100)\%$ (top) and $(1-5)\%$ (bottom) V0M classes in pp collisions at $\sqrt{s}$ = 13 TeV using PYTHIA8 (left) and AMPT (right).}
\label{fig:protonbypion}
\end{figure*}

\begin{figure*}[ht!]
\includegraphics[scale=0.29]{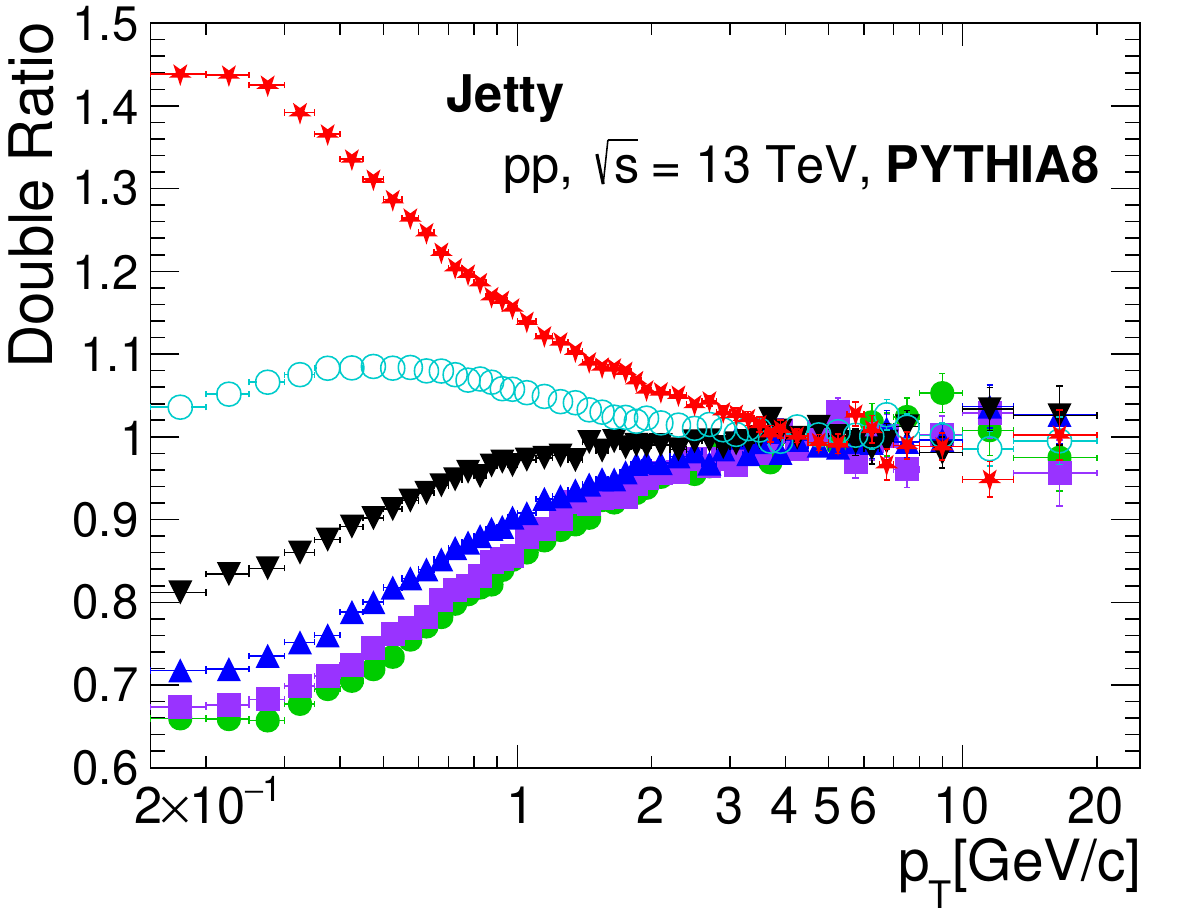}
\includegraphics[scale=0.29]{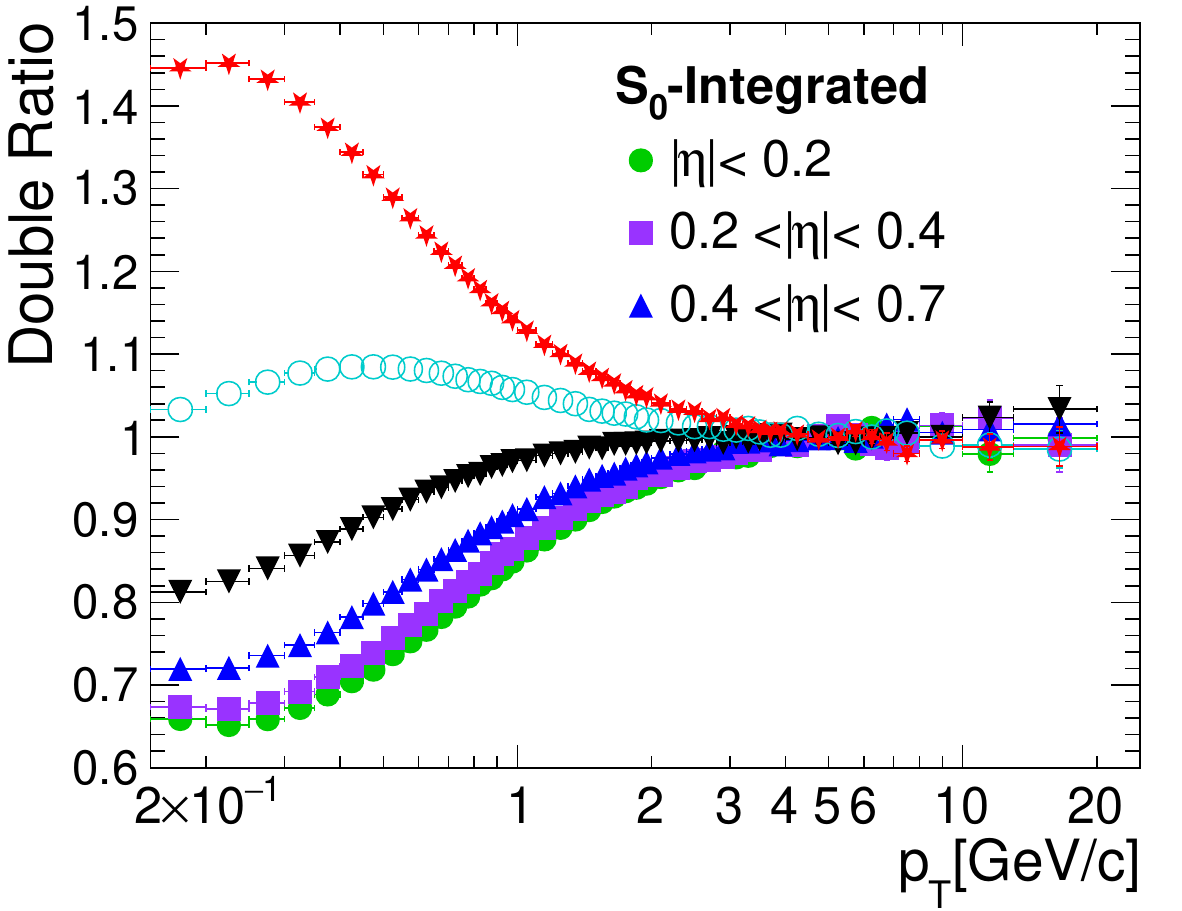}
\includegraphics[scale=0.29]{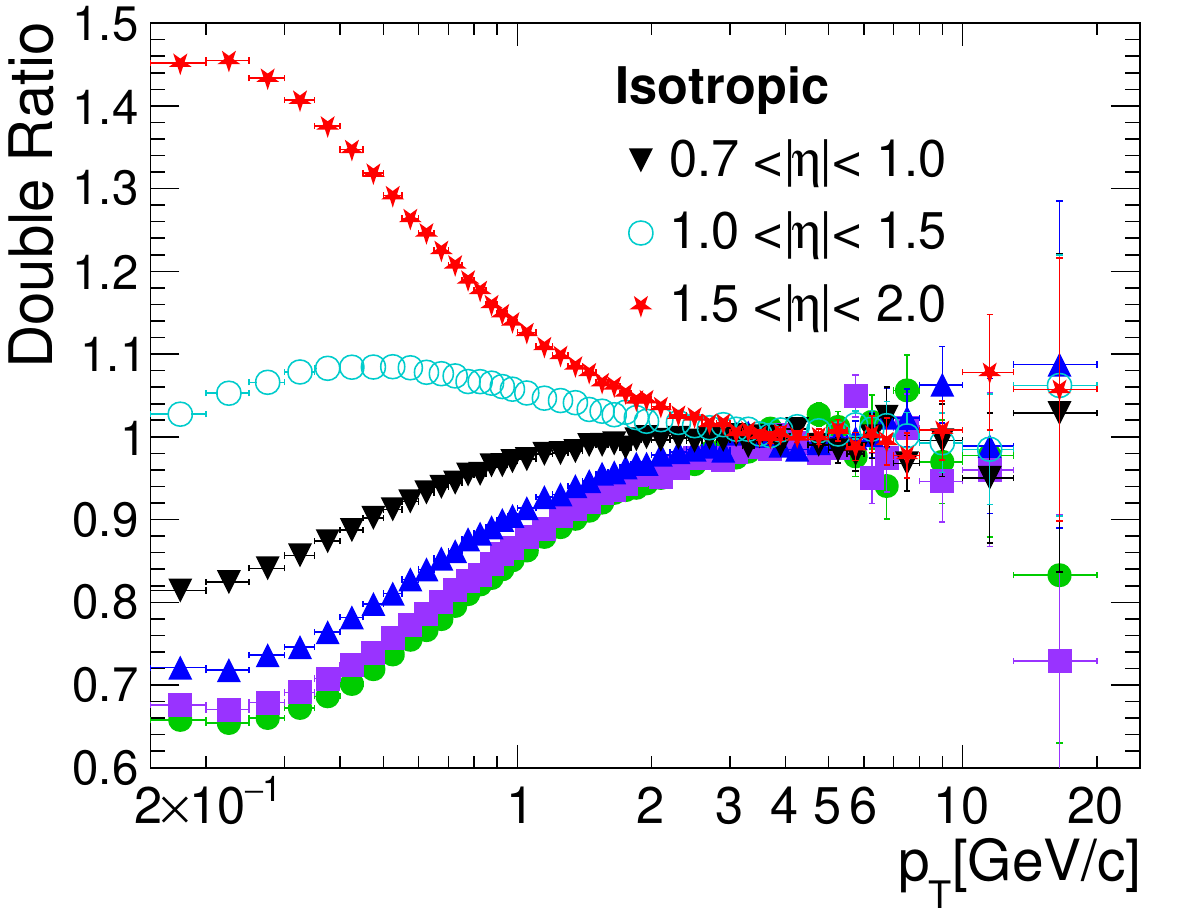}
\caption{(Color online) Proton-to-pion ratio normalized to their ratios in the pseudorapidity range, $|\eta| < 2.0$ as a function of transverse momentum for different pseudorapidity bins in pp collisions at $\sqrt{s}$ = 13 TeV using PYTHIA8 for jetty (left), spherocity-integrated (middle) and isotropic (right) events. The results are shown for $(0-100)\%$ V0M class.}
\label{fig:ppiratiotoratio}
\end{figure*}

The hydrodynamic expansion of the hot and dense partonic medium formed in the collisions of nuclear matter creates an azimuthally symmetric and radially outward flow of the particles known as the radial flow. Thus, the presence of radial flow hints towards the formation of QGP and hence, the applicability of hydrodynamics as the local thermal equilibrium of the medium is achieved. This radial flow is expected to give a larger boost to the particles with higher mass compared to the lower mass. As the massive particles gain more momentum from the flow velocity, the observed broadening in $p_{\rm T}$ spectra is expected to be different for particles from low to high masses. This mass ordering in the degree of broadening of $p_{\rm T}$ spectra signals the presence of radial flow in the system. In Ref.~\cite{Kisiel:2010xy}, authors have explicitly discussed the effect of radial flow for different particles. It has been observed that the $p_{\rm T}$ spectra of pions, in the presence of a strong radial flow, acquire a `convex' shape while protons develop a positive curvature. In the presence of radial flow, the slopes of the $p_{\rm T}$ spectra of all particles become less steep. This mass-dependent modification in the particle $p_{\rm T}$ spectra in the presence of radial flow can also be seen in the $p_{\rm T}$-dependent yield ratios of different particles. For a radially boosted system, a bump-like structure is observed in the particle yield ratios in the intermediate $p_{\rm T}$ region. This bump-like structure shifts to a higher $p_{\rm T}$ value for a more radially boosted system~\cite{ALICE:2019hno, ALICE:2013mez}. In Pb--Pb collisions at TeV energies, the proton-to-pion ratio exhibits strong enhancement (bump-like structure) in the intermediate $p_{\rm T}$ region compared to pp collisions~\cite{ALICE:2019hno, ALICE:2013mez,Preghenella:2013qsv}. %Here, the peak of the enhancement in Pb--Pb collisions shifts towards a larger $p_{\rm T}$ value compared to the pp collisions.
In addition, as one moves from central to peripheral collisions, this enhancement structure in the particle yield ratio shifts towards the lower $p_{\rm T}$ values. Radial flow is known to depend on the centrality and it decreases as one moves from the central to the peripheral collisions and hence, the shift of the bump-like structure from intermediate to low $p_{\rm T}$ is observed~\cite{ALICE:2019hno, ALICE:2013mez, Preghenella:2013qsv}.

Recently, such radial flow-like signals in the particle yield ratios are observed in high multiplicity pp collisions~\cite{ALICE:2013wgn, ALICE:2016dei}, which are otherwise taken as the baseline measurement system without QGP, for the comparison with heavy-ion collisions. Interestingly, PYTHIA8, a pQCD inspired model, can mimic the signatures of radial flow with color reconnection.  As mentioned in Ref.~\cite{OrtizVelasquez:2013ofg}, in PYTHIA8, a string that connects two partons follows the movements of the partonic endpoints, leading to a common boost of the string fragments (hadrons). In the absence of CR, when a parton is knocked out in the midrapidity, the other string end will be part of the proton moving forward, leading to a small boost. However, in the presence of CR, two partons from independent hard scatterings at similar rapidity can color reconnect and make a large transverse boost. In the presence of CR, with an increase in $N_{\rm mpi}$, the partonic collisions increase, thus the boost further increases, which consequently mimics the hydrodynamical radial flow. Although, in PYTHIA8, the source of such an observation is clearly different from hydrodynamics, PYTHIA8 mimics the features of radial flow very well. Thus we may call these features of PYTHIA8, which mimic the hydrodynamical radial flow as ``radial flow-like effects".

In this section, we try to investigate the spherocity and pseudorapidity dependence of proton to pion yield ratio in pp collisions at $\sqrt{s}$ = 13 TeV. For simplicity, we denote $(p+\bar p)/(\pi^{+}+\pi^{-})$ as $p/\pi$. The top panel in each of the four plots in Fig~\ref{fig:protonbypion} shows $p/\pi$ as a function of transverse momentum in $|\eta|<2.0$ for the jetty, $S_0$-integrated, and isotropic events using PYTHIA8 (left plots) and AMPT (right plots) while the bottom panel shows the ratio of $p/\pi$ of the jetty and isotropic events to that of the $S_0$-integrated events. The upper plots corresponds to the $(0-100)\%$ V0M multiplicity class whereas the bottom plots are for the high-multiplicity pp collisions, i.e., $(1-5)\%$ V0M class. For both the multiplicity classes, using the CR mode in PYTHIA8, the $S_0$-integrated events show a clear peak in the intermediate $p_{\rm T}$. This peak-like structure appearing in the $p/\pi$ is attributed to color reconnection in PYTHIA8~\cite{Ortiz:2016kpz, OrtizVelasquez:2013ofg}, which is analogous to the peaks in the $p/\pi$ of Pb--Pb collisions due to the hydrodynamical radial flow. Also, for both the multiplicity classes shown here, the $p/\pi$ for the isotropic events show a larger peak in the intermediate $p_{\rm T}$ as compared to the $S_0$-integrated case. This clearly hints that for isotropic events having a large value of $N_{\rm mpi}$, radial flow-like effects are enhanced. As expected, for the jetty events in $(0-100)\%$ V0M class, due to low $N_{\rm mpi}$, $p/\pi$ remains fairly flat in $p_{\rm T}$ and it means, jetty events have a very tiny contribution towards radial flow-like effects. The case becomes interesting when the jetty events in the high-multiplicity pp collisions in $(1-5)\%$ V0M class show a peak around the same $p_{\rm T}$ value as the isotropic and $S_0$-integrated events using PYTHIA8. The appearance of radial flow-like effects in the jetty events in high-multiplicity pp collisions can also be confirmed by comparing the bottom panels of both the multiplicity classes, where the ratio of the jetty to $S_0$-integrated class approaches and stays close to unity for the high-multiplicity case. This is a testimony that with spherocity selection, we can separate the rare events which show an enhanced radial flow-like behaviour. At the same time, the jetty events show the presence of radial flow-like effects but only in the high-multiplicity events. Furthermore, similar behaviour of transverse spherocity dependence of $p/\pi$ ratio can be seen in the AMPT model for the $(0-100)$\% V0M class. Although here, the bump structure of $p/\pi$ ratio in $S_0$-integrated and jetty events can not be seen, the isotropic events show a visible bump-like structure. This indicates that, even if the flow-like effects are not visible in the minimum bias events, one can use transverse spherocity to identify the events with enhanced flow-like effects. In addition, one can notice that $S_0$-integrated events for AMPT do not show a bump structure for $(1-5)\%$ V0M class in contrast to PYTHIA8, where a bump is clearly visible. However, one finds a finite bump structure for the isotropic events, which is clearly reflected in the bottom ratio plot. This indicates that in the AMPT model, although the charged particle multiplicity is not able to identify the flow-like events, the use of transverse spherocity makes a significant difference in the flow measurements.

%\textcolor{red}{spherocity and multiplicity dependence to be added after the plots are ready}
%\textcolor{teal}{Figure~\ref{fig:protonbypion} shows the proton-to-pion ratio as a function of $p_{\rm T}$ for $|\eta|<0.2$ and $1.5 <|\eta|<2.0$ ranges.
%For both the pseudorapidity ranges, isotropic events show a bigger bump structure in intermediate $p_{\rm T}$ with respect to the $S_{0}$-integrated events. Furthermore, the bump structure in isotropic events is shifted slightly towards a higher $p_{\rm T}$ value compared to $S_{0}$-integrated events. This peak-like structure is attributed to color reconnection in PYTHIA8~\cite{Ortiz:2016kpz, OrtizVelasquez:2013ofg}, which is analogous to the peaks in the $p/\pi$ ratio of Pb--Pb collisions due to the radial flow effect. This is a testimony that with spherocity selection, we are able to separate the rare events which show an enhanced-radial flow-like behavior. At the same time, the jetty events show a saturation behavior, indicating no radial flow-like effect.}

Figure~\ref{fig:ppiratiotoratio} shows the $p/\pi$ yield double ratio as a function of transverse momentum in different rapidity windows for jetty, $S_0$-integrated and isotropic events in $(0-100)\%$ V0M class. The double ratio is defined as,

\begin{eqnarray}
    {\rm Double~Ratio} = \frac{\Big((p+\bar{p})/(\pi^{+}+\pi^{-})\Big)|_{\eta \rm{-cut}}}{\Big((p+\bar{p})/(\pi^{+}+\pi^{-})\Big)|_{|\eta|<2.0}}.
\end{eqnarray}
Here, the numerator is normalized to the $p/\pi$ yield in the entire pseudorapidity range, i.e. $|\rm \eta| < 2.0$. The double $p/\pi$ ratio as a function of $p_{\rm T}$ shows a significant dependence on pseudorapidity selection at low $p_{\rm T}$ for all spherocity classes. The double ratio shows about 40\% suppression (enhancement) for $|\rm \eta| < 0.2$ (1.5 $ < |\rm \eta| < 2.0$) selection for $p_{\rm T} < 2.0$ GeV/$c$. Thereafter, the double ratio approaches a similar value for all pseudorapidity cases, and beyond $p_{\rm T} \simeq 2$ GeV/$c$, it seems to be independent of the pseudorapidity selection.

The suppression and enhancement of the double ratios observed in Fig.~\ref{fig:ppiratiotoratio} is related to the radial flow-like effects. As, in the presence of large $N_{\rm mpi}$, the radial flow-like effects are enhanced and the bump of the $p/\pi$ ratio shifts to higher-$p_{\rm T}$ as compared to a system having low radial flow-like effects, the ratio of the $p/\pi$ of the former to the latter would show a suppression at the low-$p_{\rm T}$ region. However, it is to be noted that, in the presence of regeneration effects, which are significantly high in heavy-ion collisions compared to the pp collisions, the behaviour of the double ratio for a radially boosted system may change. For a system with both regeneration and radial flow effects, we may observe a suppression at low $p_{\rm T}$ and an enhancement in the intermediate $p_{\rm T}$ region. However, in pp collisions, the regeneration effects are negligibly small. Thus, in Fig.~\ref{fig:ppiratiotoratio}, the suppression observed in the double ratio at the midrapidity indicates that the system at midrapidity exhibits stronger radial flow-like effects as compared to that at forward rapidity. This is an important finding as this behavior shows that the radial- flow-like feature, indicated by the bump-like structure in $p/\pi$ ratio at $p_{\rm T} \simeq 2-6$ GeV/$c$, is mildly affected by the pseudorapidity selection. It is also interesting to note that the double ratios show a negligible spherocity dependence for a fixed pseudorapidity selection, indicating that the spherocity preserves the capability of finding rare events with enhanced radial flow-like effects, irrespective of pseudorapidity selection.

% However, the production of soft protons compared to soft pions is higher at forward-pseudorapidity when compared to the same with mid-pseudorapidity selection.

\begin{figure}[ht!]
\includegraphics[scale=0.4]{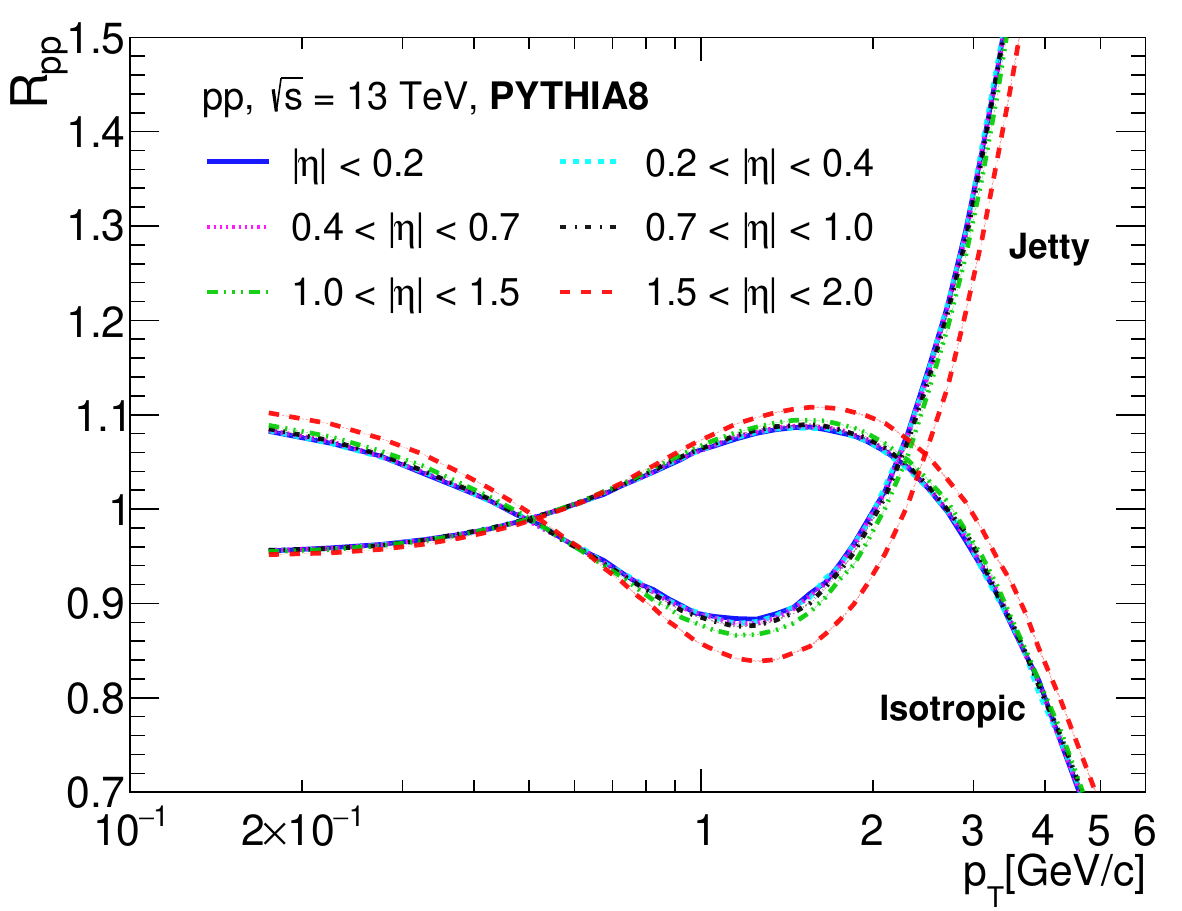}
\includegraphics[scale=0.4]{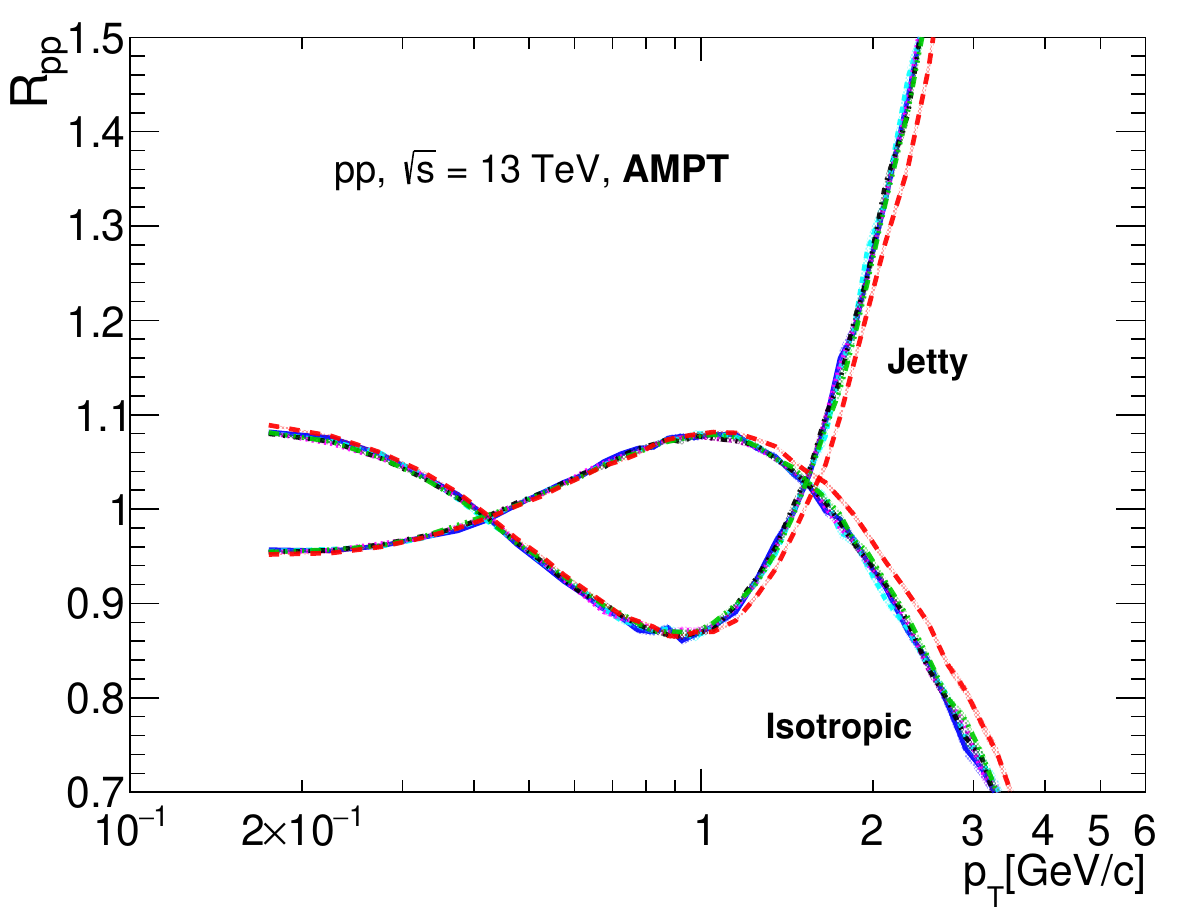}
\caption{(Color online) $R_{\rm pp}$ as a function of $p_{\rm T}$ for all charged hadrons for jetty and isotropic events within different pseudorapidity selections for $(0-100)\%$ V0M class in pp collisions at $\sqrt{s}=13$ TeV using PYTHIA8 (top) and AMPT (bottom).}
\label{fig:Rpp6eta2spheroAllCharged}
\end{figure}
\begin{figure}[ht!]
\includegraphics[scale=0.4]{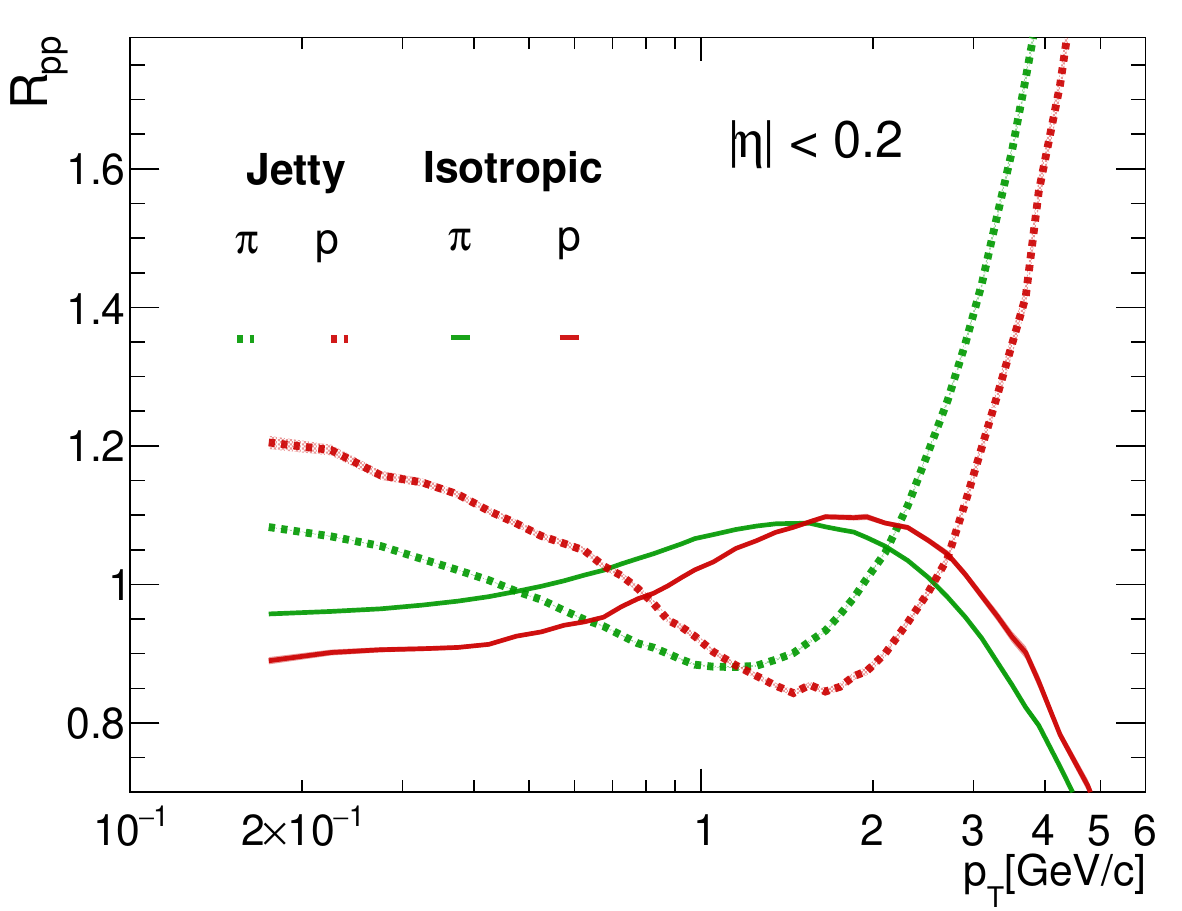}
\includegraphics[scale=0.4]{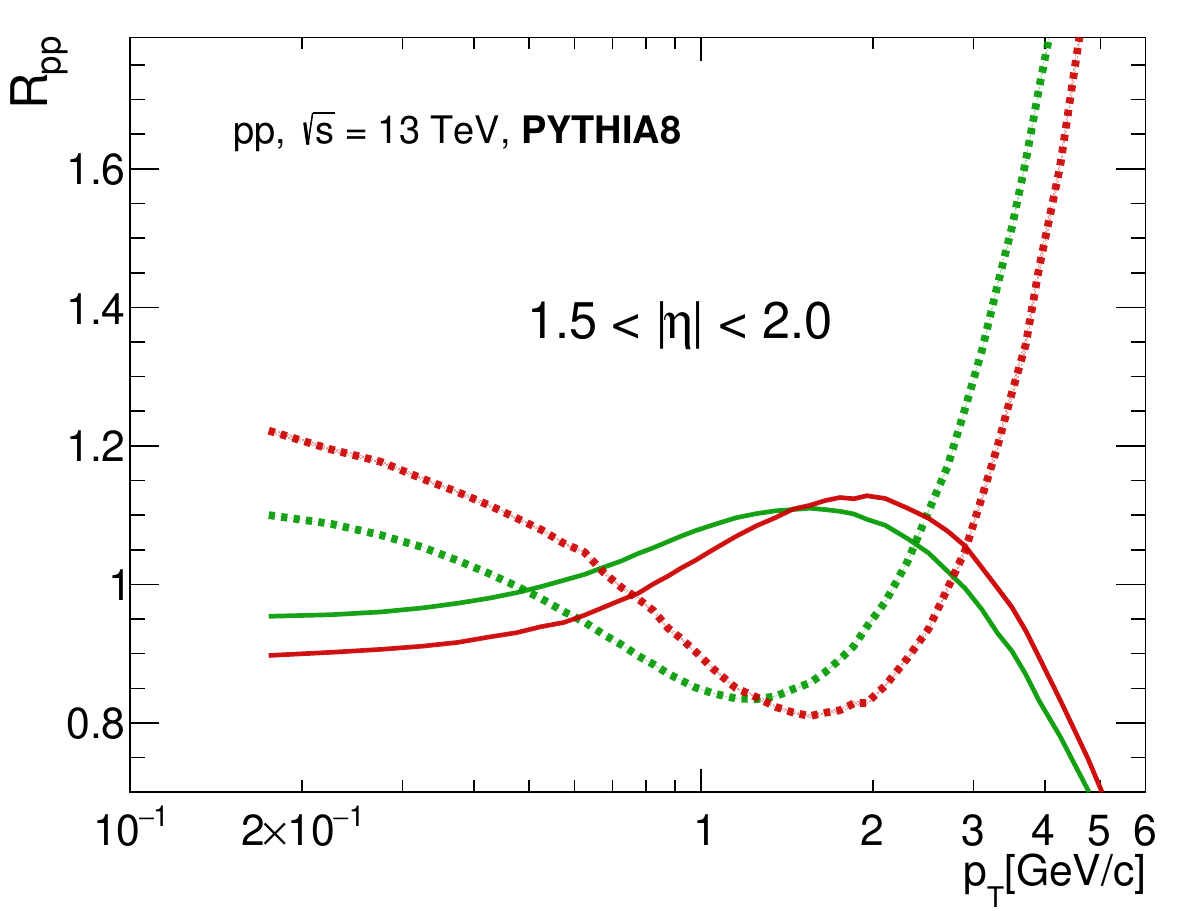}
\caption{(Color online) $R_{\rm pp}$ as a function of $p_{\rm T}$ for pions and protons for jetty and isotropic events in extreme $\eta$ classes i.e. $|\eta| < 0.2$ (top) and $1.5 < |\eta| <2.0$ (bottom) for $(0-100)\%$ V0M class in pp collisions at $\sqrt{s}=13$ TeV using PYTHIA8.}
\label{fig:RppPionProton2SpheroEta1Eta6}
\end{figure}

\subsection{Partonic modification factor \texorpdfstring{($R_{\rm{pp}}$)}{Lg}}

%In experiments, the high- $p_{\rm T}(>4~\text{GeV}/c)$ charged particle yields are observed to increase faster than the increase in the charged-particle multiplicity~\cite{ALICE:2019dfi}. This increment in the charged particle yield is smaller when we consider the lower- $p_{\rm T}$ particles, as shown in Ref.~\cite{ALICE:2019dfi}. 

It has been observed by the ALICE experiment ~\cite{ALICE:2019dfi}, that the self-normalized charged particle yields for $p_{\rm T}> 4~\text{GeV}/c$ have a faster than linear increase when studied as a function of 
average mid-pseudorapidity multiplicity. 
This non-linearity in the high-$p_{\rm T}$ particle yields is understood as a consequence of an autocorrelation bias due to the estimation of both charged particle yields and the charged particle multiplicity for event selection in the same pseudorapidity region~\cite{Weber:2018ddv}. To reduce such autocorrelation biases, the classification of events has been performed using forward pseudorapidity charged-particle multiplicity, while the observable of interest is measured in a different pseudorapidity interval~\cite{ALICE:2018pal}. Based on PYTHIA8 simulations, the charged particle multiplicity measured in forward pseudorapidity is strongly correlated to the underlying MPI activity~\cite{Ortiz:2022mfv}. However, the autocorrelation bias is still observed~\cite{ALICE:2020nkc}, which affects the observables that aim to search for medium-modification and partonic energy loss in small collision systems~\cite{ALICE:2022qxg}.  Thus, transverse spherocity, a key event shape classifier, plays an important role in reducing the sensitivity to hard processes compared to classifiers based only on forward pseudorapidity charged-particle multiplicity estimation. As this section aims to study the partonic modification factor ($R_{\rm pp}$), which is related to both radial flow and partonic energy loss, the choice of transverse spherocity as an event classifier is expected to reduce such autocorrelation biases.

Traditionally, one of the observables constructed to understand how particle production in heavy-ion collisions differs from the baseline proton-proton collisions is the nuclear modification factor, $R_{\rm AA}$. It is defined as the ratio of yield in heavy-ion collisions (A--A) to that of the yield in pp collisions normalized by the average number of binary collisions \cite{BRAHMS:2016klg, ALICE:2019hno}.
\begin{equation}
R_{\rm AA} = \frac{1}{\langle N_{\rm coll} \rangle}\frac{d^2N_{\rm AA}/d\eta dp_{\rm T}}{d^2N_{\rm pp}/d\eta dp_{\rm T}},
\label{eq:RAA}
\end{equation}
where the numerator is the $p_{\rm T}$ spectra in heavy-ion collisions while the denominator is the scaled $p_{\rm T}$ spectra in pp collisions by the total number of average binary pp collisions in a single A--A collision. In the absence of QGP medium formation, the yield in a single A--A collision is expected to be simply a linear superposition of the yield in the pp collisions times the number of binary pp collisions in a single A--A collision. This gives rise to the condition of $R_{\rm AA} = 1$, suggesting no medium effect. $R_{\rm AA} < $ 1 ($R_{\rm AA} > $ 1) for identified particles indicates a suppression (enhancement) of hadrons which can be attributed to the presence of QGP medium. In addition, $R_{\rm AA}$ can also be affected by the cold nuclear matter effects~\cite{LHCb:2021vww, Albacete:2017qng}. This includes initial state effects such as the modification of the parton distribution functions of the colliding nuclei with respect to the protons~\cite{Eskola:2016oht}.

Since in our study, the system under focus is a proton-proton collision system and as the baseline used in the definition of nuclear modification factor is also a pp collision system, there is a requirement to redefine the modification factor accordingly. In Ref.~\cite{Ortiz:2020rwg}, the authors have introduced a quantity motivated from $R_{\rm AA}$, i.e., $R_{\rm pp}$. We refer to this quantity as partonic modification factor. This is defined as the normalized ratio of the yield of all charged particles in a specific spherocity class and a particular pseudorapidity bin to their yield in the same pseudorapidity bin for $S_{\rm 0}$-integrated events. 
\begin{equation}
R_{\rm pp} = \frac{d^2N_{\rm ch}^{S_{\rm 0}}/\langle N_{\rm ch}^{S_{\rm 0}} \rangle d\eta dp_{\rm T}} {d^2N_{\rm ch}^{\rm MB}/\langle N_{\rm ch}^{\rm MB} \rangle d\eta dp_{\rm T}}\Bigg{|}_{\eta}
\label{eq:Rpp}
\end{equation}

Analogous to the $1/\langle N_{\rm coll} \rangle$ in $R_{\rm AA}$, here we normalize the yields with respect to the average charged-particle multiplicity in the corresponding spherocity event class i.e. the normalization constant used for scaling here is $\langle N_{\rm ch}^{\rm MB} \rangle / \langle N_{\rm ch}^{S_{\rm 0}} \rangle$. 

Figure~\ref{fig:Rpp6eta2spheroAllCharged} shows $R_{\rm pp}$ as a function of $p_{\rm T}$ for isotropic and jetty events in different pseudorapidity regions in pp collisions at $\sqrt{s}=13$ TeV using PYTHIA8 (top) and AMPT (bottom).
As can be seen in Fig.~\ref{fig:Rpp6eta2spheroAllCharged} using both AMPT and PYTHIA8 models, $R_{\rm pp}$ for jetty events has initially, a decreasing trend in the intermediate $p_{\rm T}$ region after which it rapidly rises towards higher $p_{\rm T}$ domains. On the other hand, $R_{\rm pp}$ for isotropic events increases first and then drops, showing a suppression trend. This behavior is very similar when one studies $R_{\rm pp}$ as a function of $p_{\rm T}$ for a different $N_{\rm mpi}$ selections as shown in Ref.~\cite{Ortiz:2020rwg}.
Here, the dip (bump) structure at the intermediate $p_{\rm T}$ followed by the enhancement (suppression) structure of $R_{\rm pp}$ at high-$p_{\rm T}$ are observed for a small (large) value of $N_{\rm mpi}$~\cite{Ortiz:2020rwg}. This dip (bump) structure at the intermediate $p_{\rm T}$ mimics a small (large) radially boosted system compared to the minimum-bias events~\cite{Ortiz:2020rwg}. 
Using transverse spherocity as an event shape classifier, as shown in Fig.~\ref{fig:Rpp6eta2spheroAllCharged}, we can extract similar features of $R_{\rm pp}$ with $N_{\rm mpi}$ selections. However, these enhancement and suppression trends in the yields for jetty and isotropic events, as shown in Fig.~\ref{fig:Rpp6eta2spheroAllCharged}, have a mild dependence on pseudorapidity. For the forward pseudorapidity range, the enhancement and suppression are steeper compared to the mid-pseudorapidity region. In addition, both PYTHIA8 and AMPT models show similar behaviour of $R_{\rm pp}$ on transverse spherocity selection. However, one finds a larger dependence of pseudorapidity selection on $R_{\rm pp}$ using PYTHIA8 as compared to AMPT.

Figure~\ref{fig:RppPionProton2SpheroEta1Eta6} shows $R_{\rm pp}$ as a function of $p_{\rm T}$ for pions and protons for jetty and isotropic events in extreme $\eta$ classes i.e. $|\eta| < 0.2$ and $1.5 < |\eta| <2.0$ in pp collisions at $\sqrt{s}=13$ TeV using PYTHIA8. A similar behavior is observed for both pions and protons as seen in Fig.~\ref{fig:Rpp6eta2spheroAllCharged}. However, the key point to note from Fig.~\ref{fig:RppPionProton2SpheroEta1Eta6} is that a significant mass-dependent shift of peak position of enhancement and suppression at intermediate-$p_{\rm T}$ is observed for jetty and isotropic events, respectively. This mass dependence of the shift in the peak positions of $R_{\rm pp}$ is attributed to the color reconnection mechanism in PYTHIA8 and is seen as radial flow-like effect, where the heavier particles are largely boosted in presence of CR as compared to lighter particles, thus shifting the peak positions to higher $p_{\rm T}$ for the heavier particles~\cite{OrtizVelasquez:2013ofg}.
%As mentioned in Ref.~\cite{OrtizVelasquez:2013ofg}, in PYTHIA8, a string that connects two partons follows the movements of the partonic endpoints, leading to a common boost of the string fragments (hadrons). In the absence of CR, when a parton is knocked out in the midrapidity, the other string end will be part of the proton moving forward, leading to a small boost. However, in the presence of CR, two partons from independent hard scatterings at midrapidity can color reconnect and make a large transverse boost. In the presence of CR, with an increase in $N_{\rm mpi}$, the partonic collisions increase, leading to enhanced flow-like effects. In addition, the CR in PYTHIA8 is observed to introduce a mass dependence of the radial flow effect~\cite{OrtizVelasquez:2013ofg}, which is also reflected in $R_{\rm pp}$, as shown in Fig.~\ref{fig:RppPionProton2SpheroEta1Eta6}.
%Furthermore, this boosting effect in PYTHIA8 in the presence of CR is similar to how hadrons are affected by flow in hydrodynamics but with a different origin~\cite{OrtizVelasquez:2013ofg}. 

\subsection{Mean transverse momentum}

\begin{figure*}[ht!]
\includegraphics[scale=0.29]{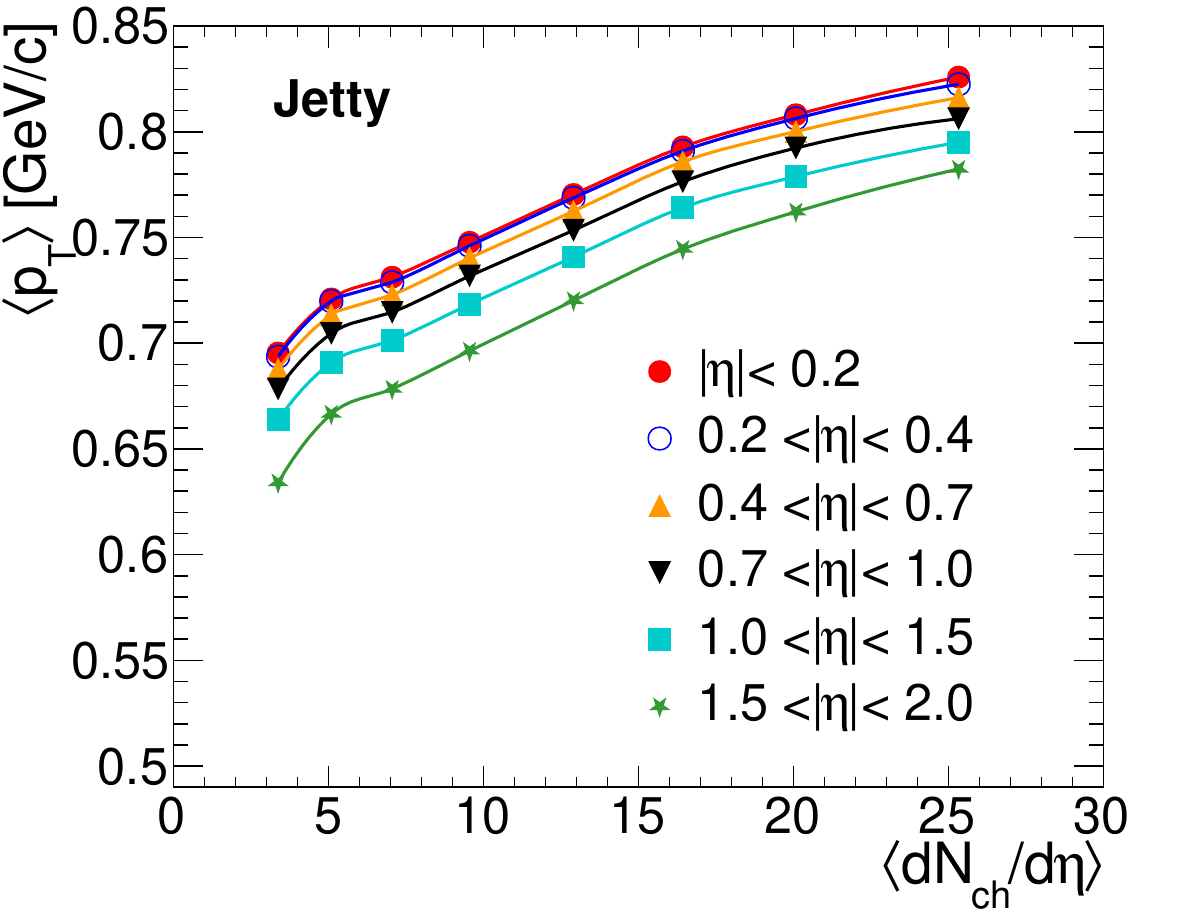}
\includegraphics[scale=0.29]{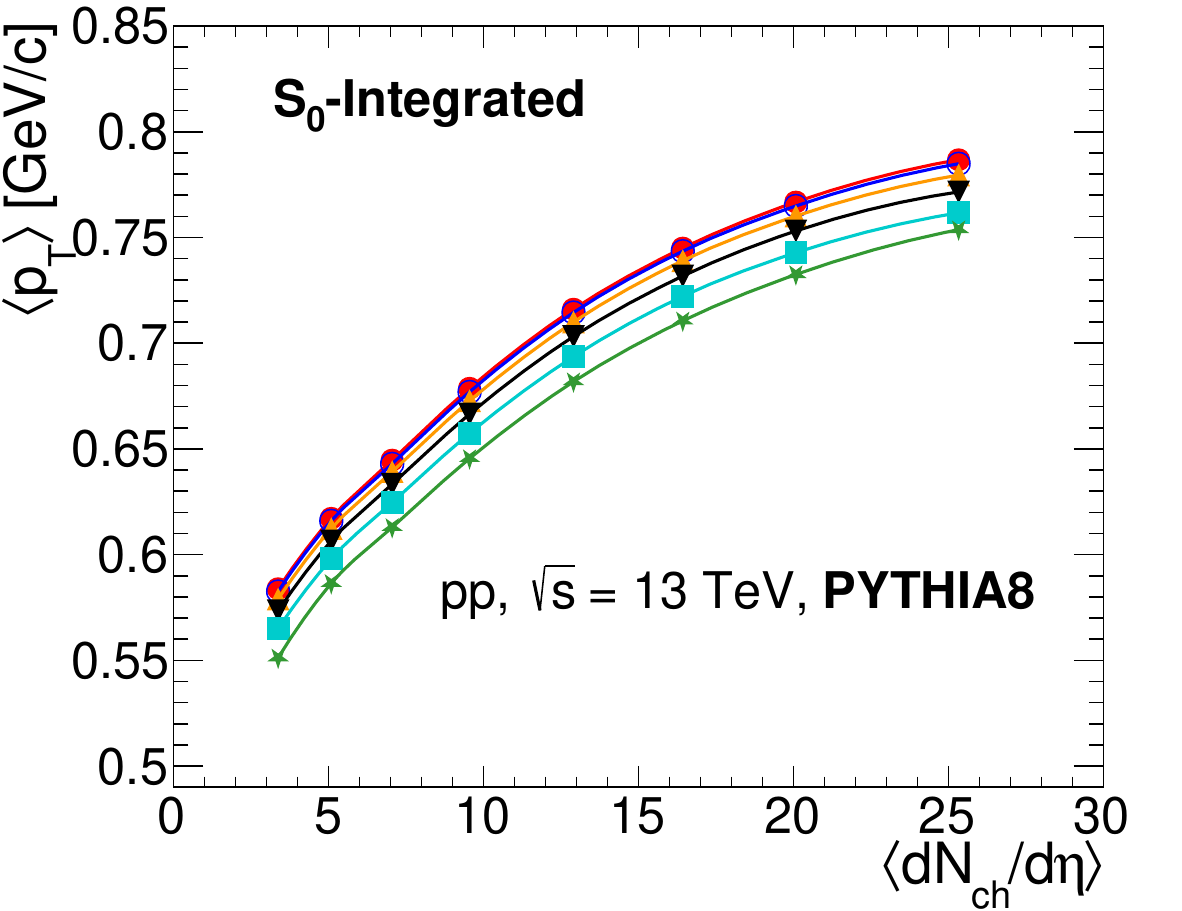}
\includegraphics[scale=0.29]{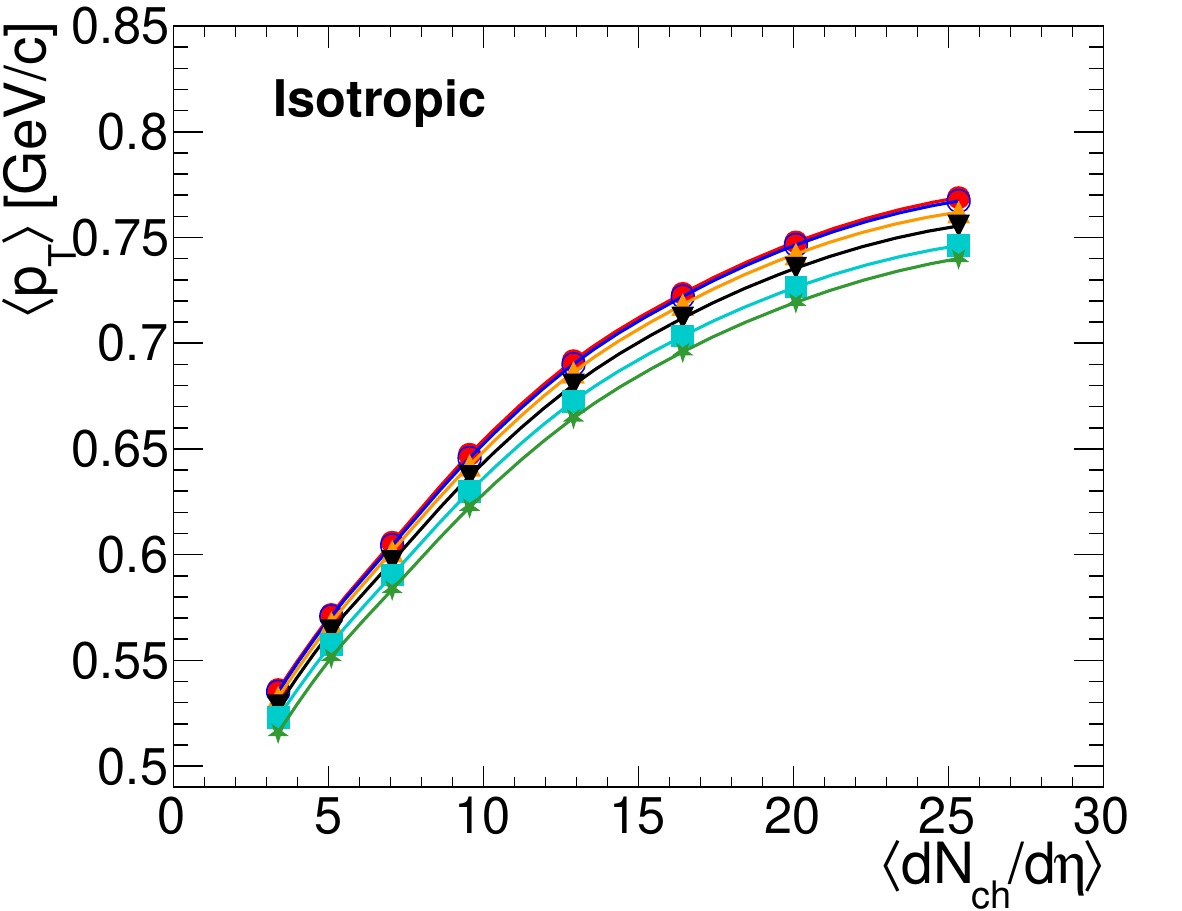}
\caption{(Color online) $\langle p_{\rm T} \rangle$ versus $\langle dN_{\rm ch}/d\eta \rangle$ as a function of pseudorapidity for jetty, $S_0$-integrated and isotropic events in pp collisions at $\sqrt{s}$ = 13 TeV using PYTHIA8.}
\label{fig:IntgMeanptNchETA}
\end{figure*}

\begin{figure*}[ht!]
\includegraphics[scale=0.29]{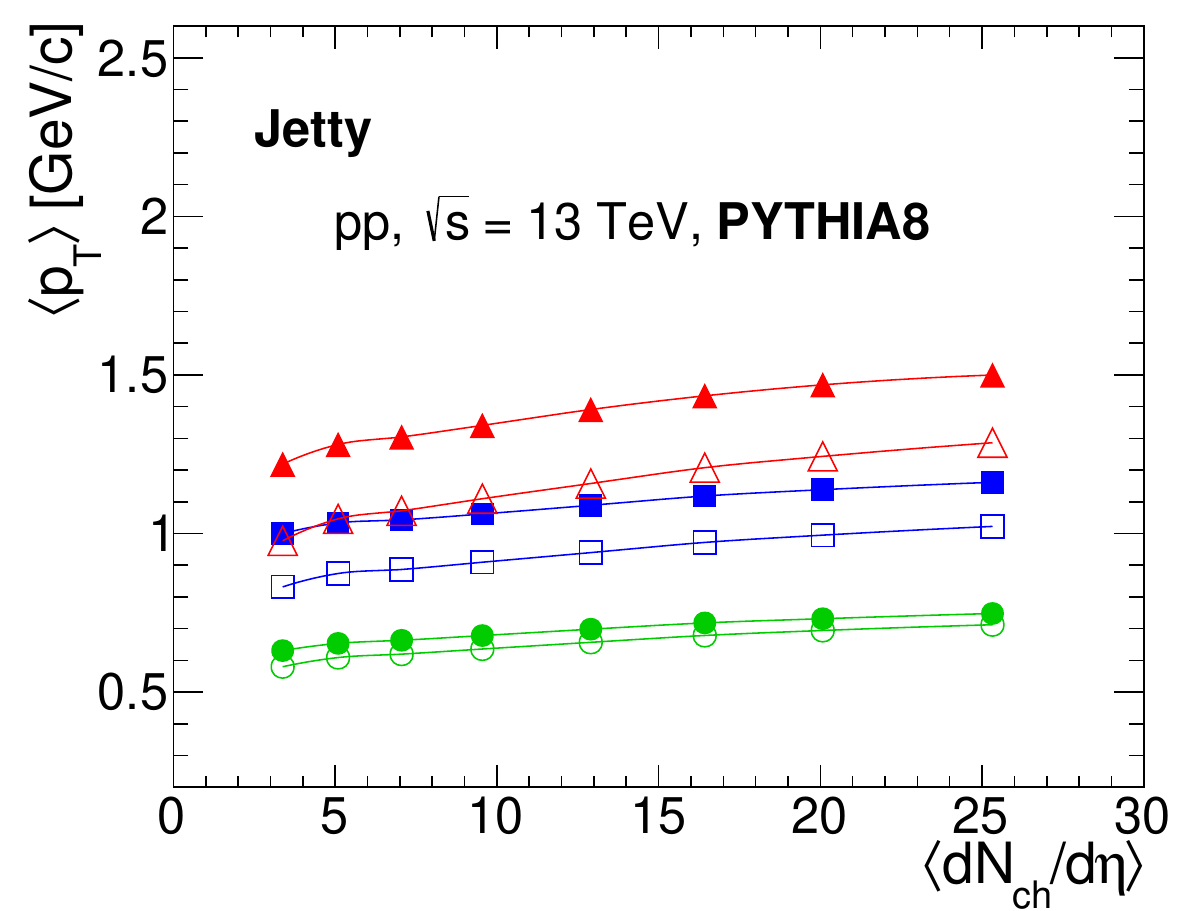}
\includegraphics[scale=0.29]{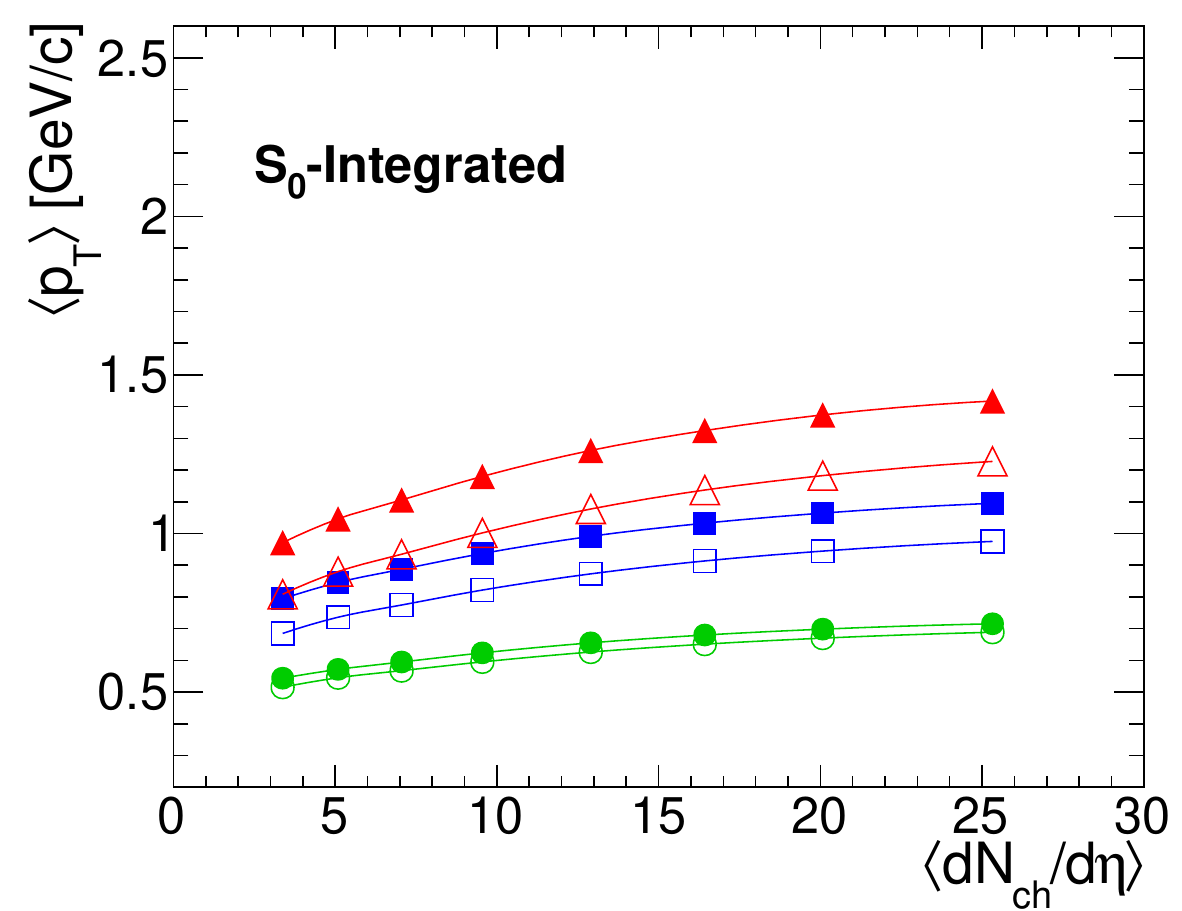}
\includegraphics[scale=0.29]{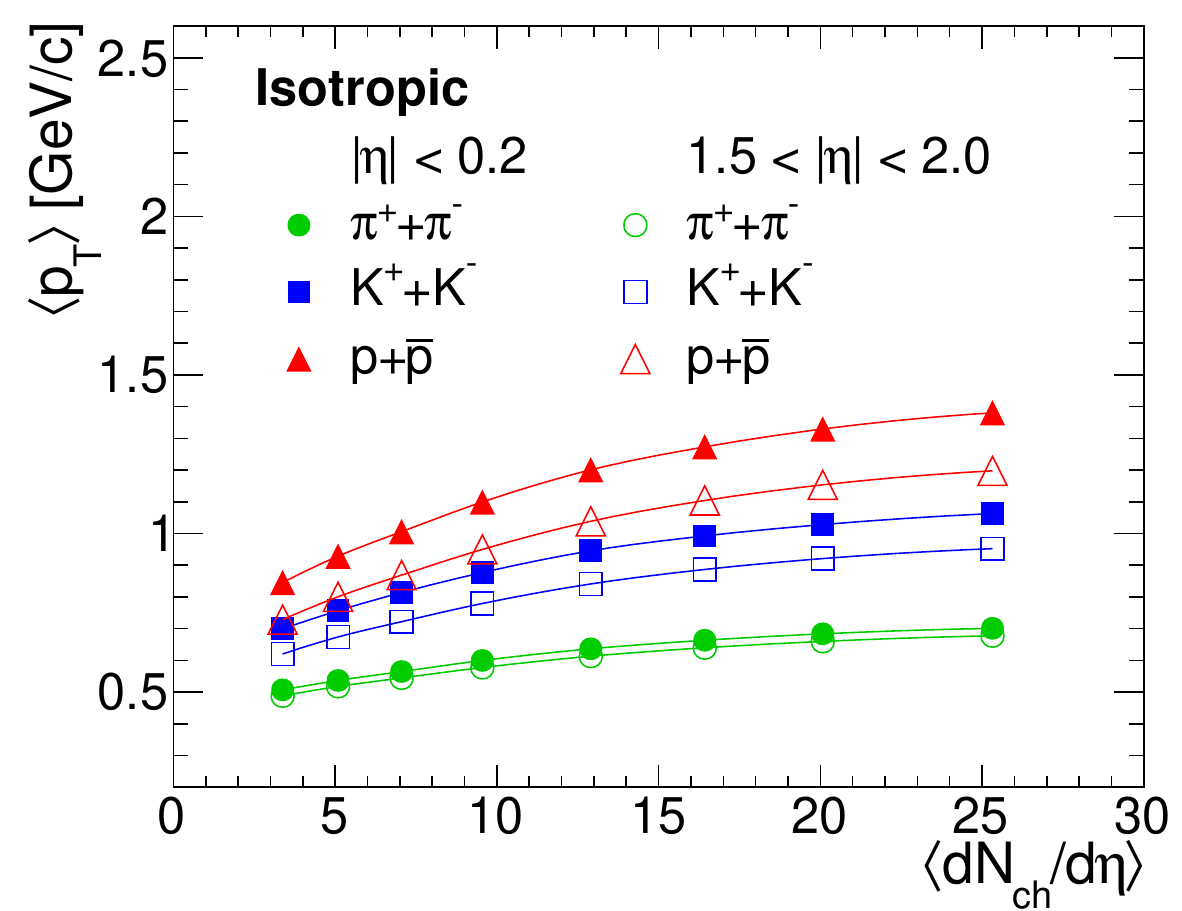}
\caption{(Color online) $\langle p_{\rm T} \rangle$ versus $\langle dN_{\rm ch}/d\eta \rangle$ for pions, kaons and protons as a function of extreme pseudorapidity classes for jetty, $S_0$-integrated and isotropic events in pp collisions at $\sqrt{s}$ = 13 TeV using PYTHIA8.}
\label{fig:IntgMeanptNchETAPID}
\end{figure*}

 %Radial flow is understood as the collective expansion of the hot and dense medium formed in relativistic collisions in a direction radially outward. The matter formed in such high-energy collisions expands and cools, and it undergoes a phase transition from the deconfined partonic state to that of a hadronic state. These produced hadrons are also carried along with the flow. As this system reaches the kinetic freeze-out, the momentum distributions of produced hadrons become fixed. In the absence of radial flow, the $p_{\rm T}$ spectra of particles in the low $p_{\rm T}$ region are expected to follow a thermal Boltzmann distribution. However, in the presence of radial flow, these exponential shapes are found to get modified \cite{Kisiel:2010xy}. This happens because radial flow induces a boost to the produced particles, which is reflected in their transverse momentum spectra. The effect is such that the high transverse momentum particles acquire a more significant boost compared to those with low transverse momentum. This leads to $p_{\rm T}$ broadening and shifting of the entire $p_{\rm T}$ spectra to higher values of transverse momenta, thus increasing the mean transverse momentum ($\langle p_{\rm T}\rangle$) of the produced hadrons.

Figure~\ref{fig:IntgMeanptNchETA} shows $\langle p_{\rm T} \rangle$ as a function of charged particle density for the jetty (left), $S_0$-integrated (middle) and isotropic (right) events estimated in different regions of pseudorapidity in pp collisions at $\sqrt{s}$ = 13 TeV from PYTHIA8. The $\langle p_{\rm T} \rangle$ is calculated in the region  $0.15 < p_{\rm T} < 10$ GeV/$c$. For all pseudorapidity selections, the $\langle p_{\rm T} \rangle$ increases with $\langle dN_{\rm ch}/d\eta\rangle$. This enhancement of $\langle p_{\rm T} \rangle$ with multiplicity follows a similar trend as observed in Pb--Pb collisions and can be attributed to higher $N_{\rm mpi}$ which leads to higher radial flow-like effects with an increase in the charged particle density~\cite{ALICE:2013rdo}. Here, one observes significant pseudorapidity dependence of $\langle p_{\rm T} \rangle$. The value of $\langle p_{\rm T} \rangle$ decreases with increased pseudorapidity. Since the mid-pseudorapidity regions, where the gluons dominate the particle production, are denser compared to forward pseudorapidities where the particle production is dominated by fragmentation and hadronic decays, the particle production is less favored at higher pseudorapidities compared to the midrapidities~\cite{Sahoo:2018osl}. Thus, due to large parton density at the mid-rapidity, the probability of color reconnection of two independent partons becomes larger, and as a consequence, larger radial flow-like effects are observed.
%Thus, the radial flow is more probable in mid-$\eta$ regions due to large particle density in the mid-pseudorapidity.
In addition, the expansion of the created partons in the longitudinal direction (forward- $\eta$) is faster than in the mid-pseudorapidity region. A faster longitudinal expansion would reduce the time for color reconnection of created partons. Thus the radial flow-like effects are diminished in forward-pseudorapidity \cite{BRAHMS:2016klg}. The $\langle p_{\rm T} \rangle$ is also found to show spherocity dependence, where $\langle p_{\rm T} \rangle$ is higher for jetty events compared to the isotropic events throughout all pseudorapidity bins. The observed higher value of $\langle p_{\rm T} \rangle$ for the jetty events is because of contribution due to the domination of jets.

To understand the pseudorapidity dependence of $\langle p_{\rm T} \rangle$ in greater detail, in Fig.~\ref{fig:IntgMeanptNchETAPID}, $\langle p_{\rm T} \rangle$ as a function of charged particle density for pions ($\pi^{+}+\pi^{-}$), kaons ($K^{+}+K^{-}$) and protons ($p+\bar{p}$) is studied for the jetty, $S_{0}$-integrated and isotropic events in the extreme pseudorapidity selections in pp collisions at $\sqrt{s}=13$ TeV using PYTHIA8. As the CR leads to a larger boost to heavier particles~\cite{OrtizVelasquez:2013ofg}, a clear mass ordering is observed where $\langle p_{\rm T} \rangle$ is more for protons than kaons and the least for pions. As expected from the trends observed in Fig.~\ref{fig:IntgMeanptNchETA}, at mid-pseudorapidity, the hadrons show a significant enhancement in $\langle p_{\rm T} \rangle$ than that of forward-pseudorapidity hadrons. This drop of $\langle p_{\rm T}\rangle$ from mid to forward pseudorapidity, with a more pronounced effect on heavier hadrons, once again reflects the diminishing nature of radial flow-like effects towards higher pseudorapidities. The observed pseudorapidity dependence of $\langle p_{\rm T}\rangle$ in Fig.~\ref{fig:IntgMeanptNchETAPID} is consistent with experimental observations in Cu--Cu collisions as shown in Ref.~\cite{BRAHMS:2016klg}.

\begin{figure*}[ht!]
\includegraphics[scale=0.42]{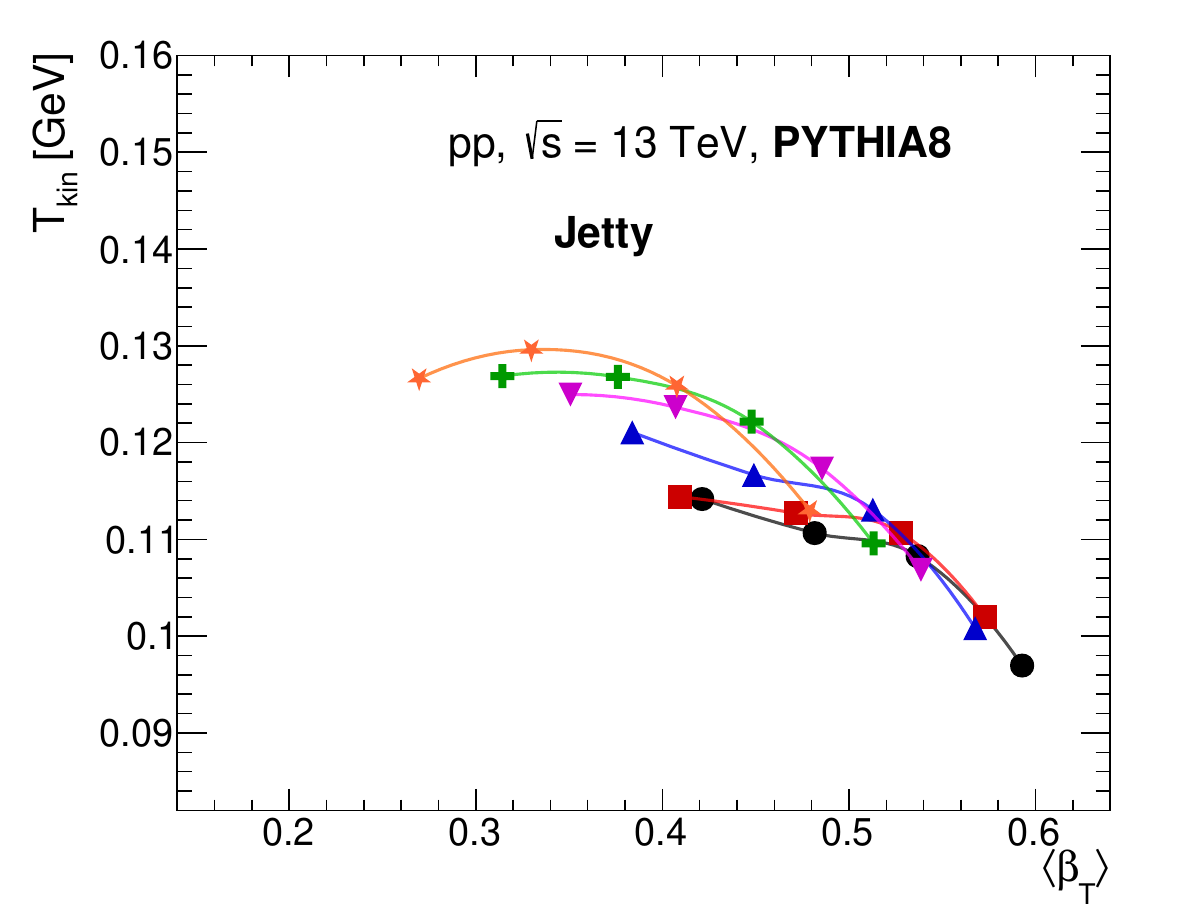}
\includegraphics[scale=0.42]{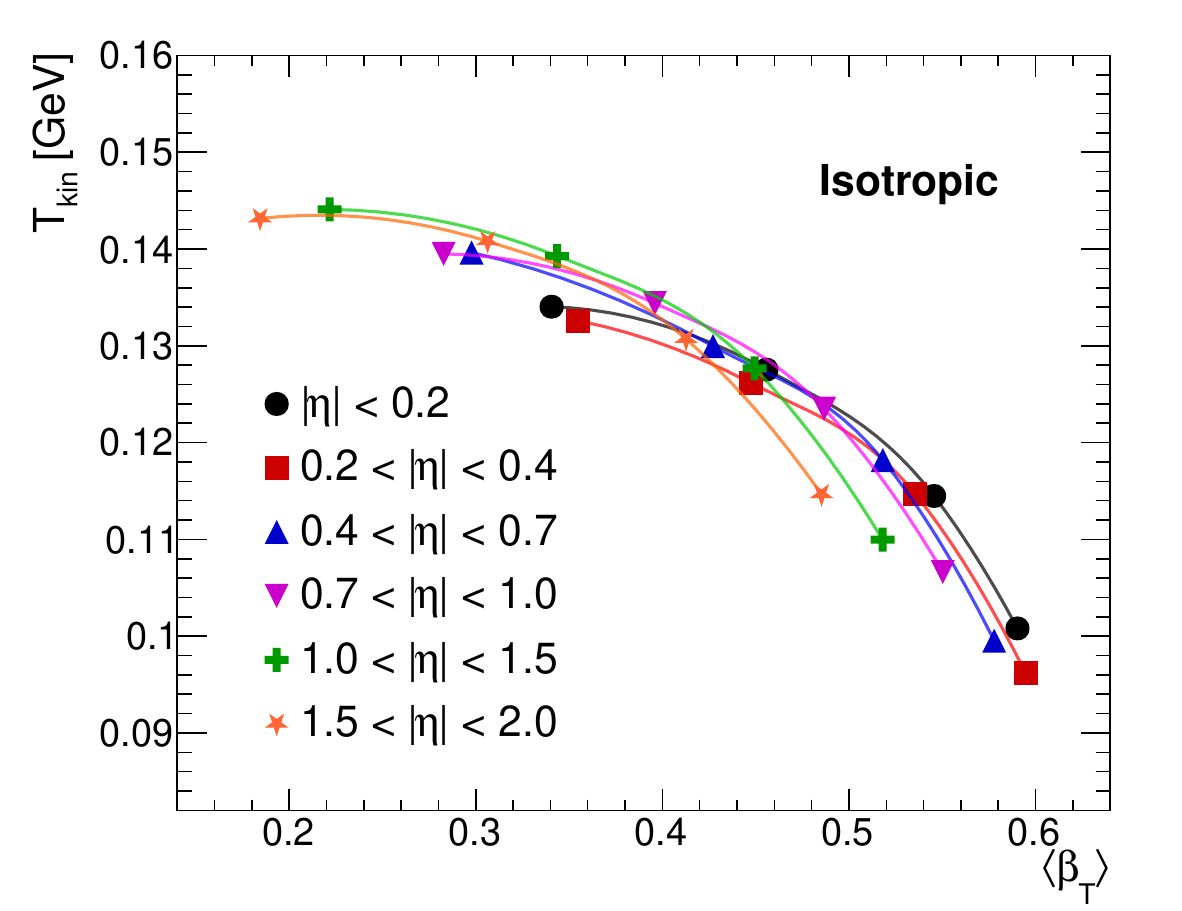}
\caption{(Color online) Kinetic freeze-out temperature versus mean transverse radial flow velocity as a function of pseudorapidity for jetty (left) and isotropic (right) events from the simultaneous blastwave fitting of the $p_{\rm T}$ spectra of protons, kaons, and pions in pp collisions at $\sqrt{s}=13$ TeV using PYTHIA8.}
\label{fig:BGBWnew}
\end{figure*}

\subsection{Kinetic freezeout parameters}
% The radially outward expansion of the hot, dense medium formed during a -- collision can be expressed using a velocity field which points radially outward from the collision centre. 

The space-time evolution in ultra-relativistic collisions involves several phases. The traditional heavy-ion picture assumes the pre-equilibrium phase, the QGP phase, the mixed phase, the chemical freeze-out, the hadron gas phase, and finally, the kinetic freeze-out phase. The effect of radial flow thus migrates from the partonic QGP phase to the hadron gas phase and survives till the kinetic freeze-out giving a boost to the produced hadrons. This results in the enhancement of the expansion velocity of the hadronic system. This expansion velocity is imprinted in the transverse momentum spectra of the produced hadrons.
As the transverse momentum spectra of identified hardons obtained using PYTHIA8 exhibit flow-like effects which are traced back to CR and MPI, it will be interesting to study the mean transverse expansion velocity ($\langle\beta_{T}\rangle$) and  kinetic freeze-out temperature ($T_{\rm kin}$). One can extract $\langle\beta_{T}\rangle$ along with $T_{\rm kin}$ by a simultaneous fitting of the Boltzmann-Gibbs blastwave (BGBW) function to the identified particles' transverse momentum spectra measured after the kinetic freeze-out. The BGBW function at midrapidity is given by the following expression \cite{BRAHMS:2016klg, Schnedermann:1993ws}:
\begin{eqnarray}
\label{boltz_blast}
\left.\frac{d^2N}{dp_{\rm T}dy}\right|_{y=0} = C p_{\rm T} m_{\rm T} \int_{0}^{R_{0}} \Bigg[ r\;D(r)\;dr\;\times \nonumber\\ K_{1}\Big(\frac{m_{T}\;\cosh\rho}{T_{\rm kin}}\Big)
I_{0}\Big(\frac{p_{\rm T}\;\sinh\rho}{T_{\rm kin}}\Big) \Bigg],
\end{eqnarray}
where $K_{1}\displaystyle\Big(\frac{m_{T}\;{\cosh}\rho}{T_{\rm kin}}\Big)$ and $I_{0}\displaystyle\Big(\frac{p_{\rm T}\;{\sinh}\rho}{T_{\rm kin}}\Big)$ are modified Bessel's functions, which are given by,
\begin{equation}
    K_{1}\Big{(}\frac{m_{\rm T}\cosh{\rho}}{T_{\rm kin}}\Big{)} = \int_{0}^{\infty} \cosh{y} \exp{\Big{(}-\frac{m_{\rm T}\cosh{y}\cosh{\rho}}{T_{\rm kin}}}\Big{)} dy,
\end{equation}
\begin{equation}
    I_{0}\Big{(}\frac{p_{\rm T}\sinh{\rho}}{T_{\rm kin}}\Big{)} = \frac{1}{2\pi}\int_{0}^{2\pi} \exp{\Big{(}\frac{p_{\rm T}\sinh{\rho}\cos{\phi}}{T_{\rm kin}}}\Big{)} d\phi,
\end{equation}
where, $\rho={\tanh}^{-1}\beta_{\rm T}$ and $\rm{\beta_{T}=\displaystyle\beta_s\xi^n}$ \cite{Schnedermann:1993ws,Huovinen:2001cy,BraunMunzinger:1994xr, Tang:2011xq}. $\rm{\beta_{T}}$ is called radial flow, $\rm{\beta_s}$ is the maximum surface velocity, $\xi=(r/R_0)$ with $r$ being the radial distance and $R_0$ the maximum radius of the source at freeze-out. $D(r)$ is the nuclear density profile; in our study, we have considered $D(r)$ = 1 for $r<R_{0}$ and $D(r)$ = 0 for $r>R_{0}$ , i.e., hard sphere profile. In this model, the particles closer to the center of the fireball are assumed to move slower than the ones at the edges. The mean transverse velocity is given by \cite{Adcox:2003nr},
 \begin{equation}
     \langle\beta_{\rm T}\rangle =\frac{\int \beta_s\xi^n\xi\;d\xi}{\int \xi\;d\xi}=\Big(\frac{2}{2+n}\Big)\beta_s.
 \end{equation}
In Eq.~\ref{boltz_blast}, the modified Bessel's function comes as a consequence of the integration from -$\infty$ to +$\infty$ over pseudorapidity $\eta$, assuming boost invariance. However, as one goes towards the forward rapidity, the condition of boost invariance is not valid. Thus, the modified Bessel's function should be replaced by an integral, i.e., $g(z)$ for $z=m_{\rm T}\cosh\rho/T_{\rm kin}$ over a finite range of $\eta$, defined as follows \cite{BRAHMS:2016klg}.

\begin{equation}
    g(z)=\int_{\eta_{\rm min}}^{\eta_{\rm max}}\;\cosh{(\eta-y)}\;e^{-z\cosh{(\eta-y)}}\;d\eta
    \label{eq:gz}
\end{equation}
Now, the substitution of Eq.~\ref{eq:gz} in Eq.~\ref{boltz_blast} modifies as follows:
\begin{eqnarray}
\label{eq:boltzblast}
\left.\frac{d^2N}{dp_{\rm T}dy}\right|_{y=0} = C p_{\rm T} m_{\rm T} \int_{0}^{R_{0}}\Bigg[ r\;D(r)\;dr\;\times \nonumber
\\g\Big(\frac{m_{T}\;\cosh\rho}{T_{\rm kin}}\Big)
I_{0}\Big(\frac{p_{\rm T}\;\sinh\rho}{T_{\rm kin}}\Big) \Bigg]
\end{eqnarray}

Using Eq.~\ref{eq:boltzblast}, one can perform the simultaneous fitting to the transverse momentum spectra of pions ($\pi^{+}+\pi^{-}$), kaons ($K^{+}+K^{-}$) and protons ($p+\bar{p}$). In Eq.~\ref{eq:gz}, the value of $y$ is chosen to be a pseudorapidity which lies exactly in the center of the polar angles corresponding to $\eta_{\rm min}$ and $\eta_{\rm max}$. We use the similar fitting range from Ref.~\cite{Cuautle:2015kra}, i.e., $0.5<p_{\rm T}<1$ GeV/$c$ for pions, $0.3<p_{\rm T}<1.5$ GeV/$c$ for kaons and $0.8<p_{\rm T}<2.0$ GeV/$c$ for protons. The values obtained from fitting for $\langle\beta_{\rm T}\rangle$ and $T_{\rm kin}$ as well as the corresponding $\chi^2/$NDF values are tabulated in Tab.~\ref{tab:betavsT}. Kinetic freeze-out temperature ($T_{\rm kin}$) versus mean transverse radial flow velocity ($\langle \beta_{\rm T} \rangle$) as a function of transverse spherocity for two extreme pseudorapidity selections are plotted in Fig.~\ref{fig:BGBWnew}. $T_{\rm kin}$ and $\langle \beta_{\rm T} \rangle$ are found to be multiplicity dependent. Higher multiplicity classes show increased $\langle \beta_{\rm T} \rangle$ and decreased $T_{\rm kin}$ which ought to be due to the longer time it would require to reach the freeze-out~\cite{Prasad:2021bdq}. In addition, for a given class of transverse spherocity, because of the higher particle density at the mid-pseudorapidity, the particles at mid-pseudorapidity show a higher value of $\langle\beta_{\rm T}\rangle$ and lower value of $T_{\rm kin}$ compared to the particles at the forward-pseudorapidity. Furthermore, a mild spherocity dependence is observed for both $\langle \beta_{\rm T} \rangle$ and $T_{\rm kin}$. For the high multiplicity class, the results are consistent with Ref.~\cite{Prasad:2021bdq} for Pb--Pb collisions at $\sqrt{s_{\rm NN}}$ = 5.02 TeV, i.e., the isotropic events possess higher value of $\langle\beta_{\rm T}\rangle$ and lower $T_{\rm kin}$ compared to the jetty events. However, as one moves towards the lower multiplicity class, due to dominating effects from the initial hard scatterings and jet fragmentations, the jetty events acquire higher values of $\langle\beta_{\rm T}\rangle$ and lower $T_{\rm kin}$.

 \begin{table*}[ht!]
\begin{tabular}{|c||c||c|c|c|c|c|c|}
\hline
\multirow{2}{*}{\bf{$\eta$ class}} & \multirow{2}{*}{\bf{V0M Percentile}} & \multicolumn{3}{c|}{\bf{Isotropic}} & \multicolumn{3}{c|}{\bf{Jetty}} \\\cline{3-8}
                    & & $\langle\beta_{\rm T}\rangle$  &$T_{\rm kin}$ [GeV]  & $\chi^2/\rm{NDF}$  & $\langle\beta_{\rm T}\rangle$  & $T_{\rm kin}$ [GeV]  & $\chi^2/\rm{NDF}$ \\\hline\hline

\multirow{4}{*}{\bf{$|\eta|< 0.2$}} 

                      &0-1 & 0.590 $\pm$ 0.010 & 0.101 $\pm$ 0.007  &0.547 &  0.593 $\pm$ 0.008 & 0.097 $\pm$ 0.005  &0.833 \\ \cline{2-8}                                 
                     &10-20 & 0.546 $\pm$ 0.013  & 0.114 $\pm$ 0.006  &0.674 &  0.537 $\pm$ 0.009 & 0.108 $\pm$ 0.004 & 1.049     \\ \cline{2-8}
                     &30-40 & 0.456 $\pm$ 0.024  & 0.128 $\pm$ 0.008  &1.260  & 0.482 $\pm$ 0.014  & 0.111 $\pm$ 0.006  & 0.833 \\ \cline{2-8} 
                     &50-70 & 0.341 $\pm$ 0.042 &  0.134 $\pm$ 0.009  &1.640 & 0.421 $\pm$ 0.018  &  0.114 $\pm$ 0.006  &1.196 \\ \hline \hline

\multirow{4}{*}{\bf{$0.2<|\eta|<0.4$}} 
                   
                    &0-1 & 0.595 $\pm$ 0.010  &  0.096 $\pm$ 0.007 &0.528 &   0.573 $\pm$ 0.012  &  0.102 $\pm$ 0.009  & 0.539   \\ \cline{2-8} 
                    &10-20 & 0.535 $\pm$ 0.013 & 0.115 $\pm$ 0.006  &0.674 & 0.528 $\pm$ 0.010  & 0.111 $\pm$ 0.006 &1.127                              \\ \cline{2-8}
                    &30-40   &0.447 $\pm$ 0.024    &0.126 $\pm$ 0.008 &0.909 &  0.472 $\pm$ 0.012  & 0.113 $\pm$ 0.005 & 1.321                              \\\cline{2-8}
                    &50-70 &  0.355 $\pm$ 0.038  &  0.133 $\pm$ 0.009  & 1.678 & 0.409 $\pm$ 0.017 & 0.114 $\pm$ 0.006  &2.059 \\ \hline\hline

\multirow{4}{*}{\bf{$0.4<|\eta|<0.7$}} 
                    
                    &0-1 & 0.578 $\pm$ 0.013  &0.100 $\pm$ 0.008 &0.307  & 0.568 $\pm$ 0.012 & 0.101 $\pm$ 0.008  & 0.475    \\ \cline{2-8}  
                    &10-20 & 0.518 $\pm$ 0.014 &  0.118 $\pm$ 0.007  &0.525  &  0.513 $\pm$ 0.014  &  0.113 $\pm$ 0.007 & 0.546  \\ \cline{2-8}
                    &30-40  & 0.427 $\pm$ 0.027   & 0.130 $\pm$ 0.009   & 1.082   & 0.449 $\pm$ 0.017     & 0.117 $\pm$ 0.007      &  0.848                                 \\ \cline{2-8}
                    &50-70 & 0.298 $\pm$ 0.040    &0.140 $\pm$ 0.008     &1.463    & 0.384 $\pm$ 0.019    & 0.121 $\pm$ 0.006      &  0.880                                   \\ \hline\hline

\multirow{4}{*}{\bf{$0.7<|\eta|<1.0$}} 

                         &0-1 & 0.550 $\pm$ 0.015  & 0.107 $\pm$ 0.009 & 0.275 & 0.539 $\pm$ 0.014 & 0.107 $\pm$ 0.008 &  0.664    \\ \cline{2-8}  
                    &10-20 & 0.487 $\pm$ 0.015 &  0.124 $\pm$ 0.006  & 0.568 &  0.486 $\pm$ 0.015  & 0.117 $\pm$ 0.007 & 0.459                 \\ \cline{2-8}
                    &30-40   & 0.396 $\pm$ 0.028    & 0.134 $\pm$ 0.008    &0.851    & 0.407 $\pm$ 0.018    & 0.124 $\pm$ 0.006      &  0.631    \\ \cline{2-8}
                    &50-70 & 0.283 $\pm$ 0.037     &  0.140 $\pm$ 0.007    & 0.731  & 0.351 $\pm$ 0.023   &  0.125 $\pm$ 0.007     &0.559 \\ \hline\hline

 \multirow{4}{*}{\bf{$1.0<|\eta|<1.5$}} &0-1 &0.222 $\pm$ 0.021    & 0.144 $\pm$ 0.004    & 1.936  &  0.314 $\pm$ 0.017    &0.127 $\pm$ 0.005       & 0.733                            \\ \cline{2-8}

                         &10-20   & 0.344 $\pm$ 0.019  & 0.139 $\pm$ 0.005    &1.236    & 0.376 $\pm$ 0.014    &  0.127 $\pm$ 0.005    &  0.900                                 \\ \cline{2-8}
                    &30-40 & 0.449 $\pm$ 0.012 & 0.128 $\pm$ 0.005   & 0.782  & 0.448 $\pm$ 0.012  &0.122 $\pm$ 0.005    & 0.696                                \\ \cline{2-8}
                    &50-70 & 0.518 $\pm$ 0.009  &0.110 $\pm$ 0.005  & 0.527 & 0.513 $\pm$ 0.008  &0.110 $\pm$ 0.004   &  0.542                 \\ \hline\hline

\multirow{4}{*}{\bf{$1.5<|\eta|<2.0$}}       &0-1 & 0.485 $\pm$ 0.010  &0.115 $\pm$ 0.005  & 0.515 & 0.479 $\pm$ 0.010 & 0.113 $\pm$ 0.005  & 0.616       \\ \cline{2-8}

                    &10-20 &0.413 $\pm$ 0.013  & 0.131 $\pm$ 0.005   & 0.811  & 0.408 $\pm$ 0.012  & 0.126 $\pm$ 0.005   & 0.745                                \\ \cline{2-8}
                    &30-40   & 0.306 $\pm$ 0.017  & 0.141 $\pm$ 0.004    & 1.322   &0.330 $\pm$ 0.014    & 0.130 $\pm$ 0.004     & 0.856                                  \\ \cline{2-8}
                                    
                     &50-70 &0.184 $\pm$ 0.026   & 0.143 $\pm$ 0.004    & 1.941  & 0.270 $\pm$ 0.014     &  0.127 $\pm$ 0.003     & 1.327 \\ \hline\hline     
\end{tabular}
\caption{\label{tab:betavsT} Kinetic freeze-out temperature ($T_{\rm kin}$), mean transverse radial flow velocity ($\langle\beta_{\rm T}\rangle$) and $\chi^{2}$/NDF values obtained from a simultaneous fit of identified charged particles' $p_{\rm T}$ spectra with Boltzmann-Gibbs blastwave function as a function of multiplicity, pseudorapidity and transverse spherocity classes in pp collisions at $\sqrt{s}$ = 13 TeV from PYTHIA8.}
\end{table*}
\begin{figure}[ht!]
\includegraphics[scale=0.35]{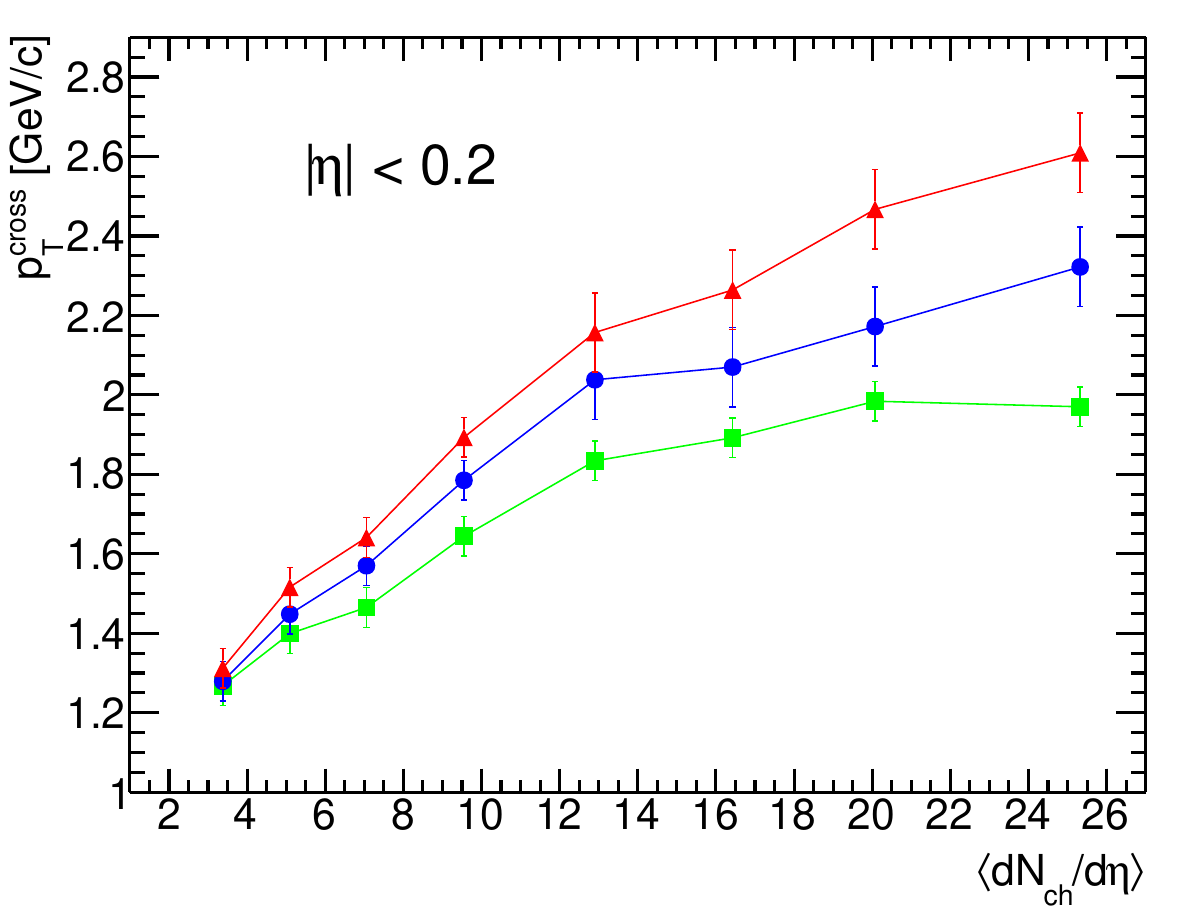}
\includegraphics[scale=0.35]{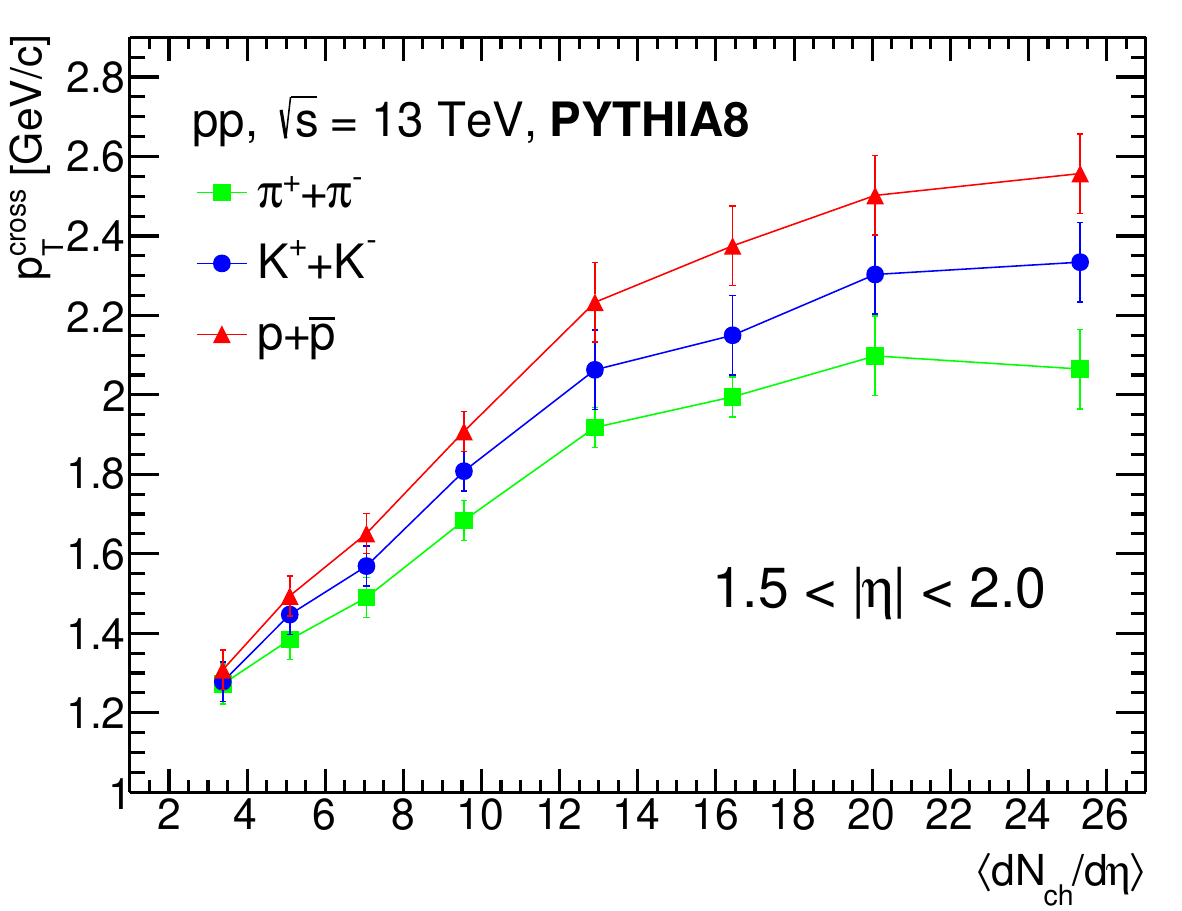}
\caption{(Color online) The crossing of the ratio of $p_{\rm T}$ spectra for isotropic and jetty events ($p_{\rm T}^{\rm cross}$) as a function of mean charged particle multiplicity density ($\langle dN_{\rm ch}/d\eta\rangle$) for pions, kaons and protons in extreme pseudorapidity bins for pp collisions at $\sqrt{s}=13$ TeV from PYTHIA8.}
\label{fig:crossingpoints}
\end{figure}

\subsection{Transverse momentum crossing points}
 
 The $p_{\rm T}$ values where the $p_{\rm T}$ spectra for jetty events start to dominate over the isotropic events are referred to as the $p_{\rm T}$ crossing points ($p_{\rm T}^{\rm cross}$). These crossing points suggest that the production of hadrons is dominated by isotropic events in the low $p_{\rm T}$ and jetty events in the high $p_{\rm T}$ region \cite{Khuntia:2018qox}. Figure~\ref{fig:crossingpoints} shows $p_{\rm T}^{\rm cross}$ for two extreme pseudorapidity selections studied as a function of mean charged particle multiplicity density ($\langle dN_{\rm ch}/d\eta\rangle$) for pions, kaons and protons in pp collisions at $\sqrt{s}=13$ TeV using PYTHIA8. As one can observe, the crossing of $p_{\rm T}$ spectra of isotropic and jetty events has a clear multiplicity dependence, and the crossing point shifts towards higher $p_{\rm T}$ with increasing multiplicity. This would imply that the soft particle production in high multiplicity collisions lasts to higher values in $p_{\rm T}$ compared to low multiplicity events. %..\textcolor{violet}{More radial flow boost in high multiplicity case as compared to low multiplicity events.} 

In Fig.~\ref{fig:crossingpoints}, we also see a mass dependence where shifting of $p_{\rm T}$ crossing points to higher $p_{\rm T}$ is more for protons than kaons followed by pions, and this could be attributed to a flow-like behaviour caused by CR. This mass ordering is observed in all pseudorapidity windows, with the mass dependence becoming clearer only for high multiplicity classes. As discussed in earlier sections, CR with a larger $N_{\rm mpi}$ causes the transverse momentum of heavier particles to suffer a higher degree of broadening compared to the lighter particles. This effect shifts the crossing point to higher values for massive particles compared to the light mass particles. Due to low $N_{\rm mpi}$ for a lower multiplicity class, the broadening of $p_{\rm T}$-spectra is small and the crossing points for different particles are indistinguishable.  As one moves towards the higher multiplicity class, the radial flow-like effects from CR shifts the crossing point to a higher $p_{\rm T}$ value for heavier particles, leading to a clear distinction between the crossing points of different masses. However, one does not observe any pseudorapidity dependence on the crossing points within the uncertainties.

\section{Summary}
\label{summary}

This study aims to explore the pseudorapidity and transverse spherocity dependence of radial flow-like effects in pp collisions at $\sqrt{s}= 13~\rm TeV$ using AMPT and a pQCD-inspired model PYTHIA8, which reproduces the experimental spectra and flow-like signatures observed in experiments by implementing CR and MPI. Unlike the earlier definitions of spherocity in the pseudorapidity range $|\eta| < 0.8$, here, we redefine spherocity in the $|\eta| < 2.0$ region. With this modification in spherocity, we perform the differential studies of various observables that are known to be sensitive to the radial flow-like effects as a function of transverse spherocity, pseudorapidity (upto $|\eta| < 2.0$) and charged-particle multiplicity wherever possible. These observables include $\langle p_{\rm T}\rangle$, particle ratios such as $p/\pi$, kinetic freezeout parameters such as $T_{\rm kin}$ and $\langle\beta_{\rm T}\rangle$, partonic modification factor ($R_{\rm pp}$) and the transverse momentum crossing points between the jetty and isotropic events. For selected observables, a comparative study using AMPT, a model based on kinetic-theory, is also performed. The findings are summarised below.

\begin{enumerate}
 %   \item The spherocity defined in the pseudorapidity region, $|\eta|<2.0$, has its peak shifted more towards the isotropic limit than the spherocity defined in the $|\eta|<0.8$ region.
 \item The mean number of multipartonic interactions has a direct correlation with the transverse spherocity. 
    %\item The increase in $K/\pi$ ratio with $p_{\rm T}$, due to larger strangeness production, is more for isotropic events than jetty events at higher $p_{\rm T}$, which can be attributed to the more partonic interactions in the former.
\item The $p/\pi$ ratio as a function of $p_{\rm T}$ shows a peak-like structure at around 3~GeV/$c$ where the spherocity dependence also becomes clearly distinct. Peak exhibited by isotropic events are more prominent than jetty events due to large $N_{\rm mpi}$, which shows that the flow-like effects are enhanced in isotropic events. 

% \item At intermediate $p_{\rm T}$, the $p/\pi$ ratios are slightly shifted to higher $p_{\rm T}$ value for hadrons at the mid-pseudorapidity as compared to forward-pseudorapidity indicating a slightly higher radial flow like effects at the mid-pseudorapidity region.

\item At low to intermediate $p_{\rm T}$, the suppression of the double ratio for the particles at mid-pseudorapidity signifies the presence of a large MPI activity as compared to the particles at the higher pseudorapidity regions, where an enhancement in the double ratio is observed.

\item Isotropic and jetty events show suppression and enhancement trends in partonic modification factor ($R_{\rm pp}$) at higher $p_{\rm T}$ values, respectively. These effects are mildly enhanced at the central pseudorapidity region. $R_{\rm pp}$ shows a particle mass dependence, where the effects are enhanced for protons than pions.

\item The isotropic events show lower $\langle p_{\rm T}\rangle$ than the jetty events due to large contributions from jets in the jetty events. $\langle p_{\rm T}\rangle$ increases with an increase in the charged particle multiplicity and towards the central pseudorapidity regions, indicating the effect from a large MPI activity leading to an enhanced radial flow-like scenario. The mass dependence of $\langle p_{\rm T}\rangle$ for different rapidity regions indicates that the heavier mass particles receive a higher boost from CR compared to the lighter particles.

\item The kinetic freeze-out parameters viz. $T_{\rm kin}$ and $\langle \beta_{\rm T}\rangle$ have clear multiplicity dependence, where an increase in multiplicity is associated with an enhanced flow velocity and a decreased kinetic freezeout temperature. Particles at central pseudorapidity regions possess higher $\langle \beta_{\rm T}\rangle$ and lower $T_{\rm kin}$ compared to the forward rapidity regions.

\item The transverse momentum crossing points have strong multiplicity dependence with $p_{\rm T}^{\rm cross}$ shifting to higher values of $p_{\rm T}$ with increasing multiplicity. This indicates that the soft particle production lasts to higher values of $p_{\rm T}$ in high multiplicity events.

\item The dependence of $p_{\rm T}^{\rm cross}$ on $\langle dN_{\rm ch}/ d\eta \rangle$ also shows a mass-ordering at all pseudorapidity bins and this becomes clearer at higher multiplicities. The observed shift in $p_{\rm T}^{\rm cross}$ with multiplicity is higher for protons, followed by kaons and pions, which might also be pointing at the flow-like behaviour and can be attributed to CR.
\end{enumerate}

In summary, the isotropic events are found to show significantly larger flow-like effects than the jetty events. At low multiplicities, several observables related to radial flow-like effects are prone to contributions from hard processes that are not part of MPI. In addition, the heavier particles are observed to gain a larger boost than the lighter particles from CR, which is consistent with hydrodynamic simulations. Furthermore, the pseudorapidity dependence of the radial flow-like effects, which is new in pp collisions, is explored by studying several observables. Experimentally observed radial flow signatures at midrapidities in high multiplicity pp collisions persist in the higher rapidities but with a reduced magnitude. This indicates that MPI activity has more significant effect on the hadrons produced at the mid-rapidity as compared to forward rapidity. The rapidity dependent effects are similar to the experimental results for Cu--Cu collisions~\cite{BRAHMS:2016klg}. As the ALICE detector at CERN has a planned upgrade to extend particle identification to $|\eta|<4$, this study can motivate the experimentalists to cross-verify the findings with a clear physics goal in mind. In addition, comparing QGP signatures in pp collisions with those of heavy-ion collisions as a function of pseudorapidity may help the collider physics community understand the production of QGP droplets in pp collisions from a new angle.

\section*{Data Availability Statement}  
This is a phenomenological paper with Monte-Carlo event generator data. In case the data are required by any of the readers, we shall provide the same upon request to the corresponding author.

\section*{Acknowledgement}
A.M.K.R. acknowledges the doctoral fellowships from the DST INSPIRE program of the Government of India. 
S.P. acknowledges the University Grants Commission (UGC), Government of India. The authors gratefully acknowledge the DAE-DST, Government of India funding under the mega-science project “Indian participation in the ALICE experiment at CERN” bearing Project No. SR/MF/PS-02/2021-IITI(E-37123).

\section*{Appendix}
\begin{figure}[ht!]
\includegraphics[scale=0.42]{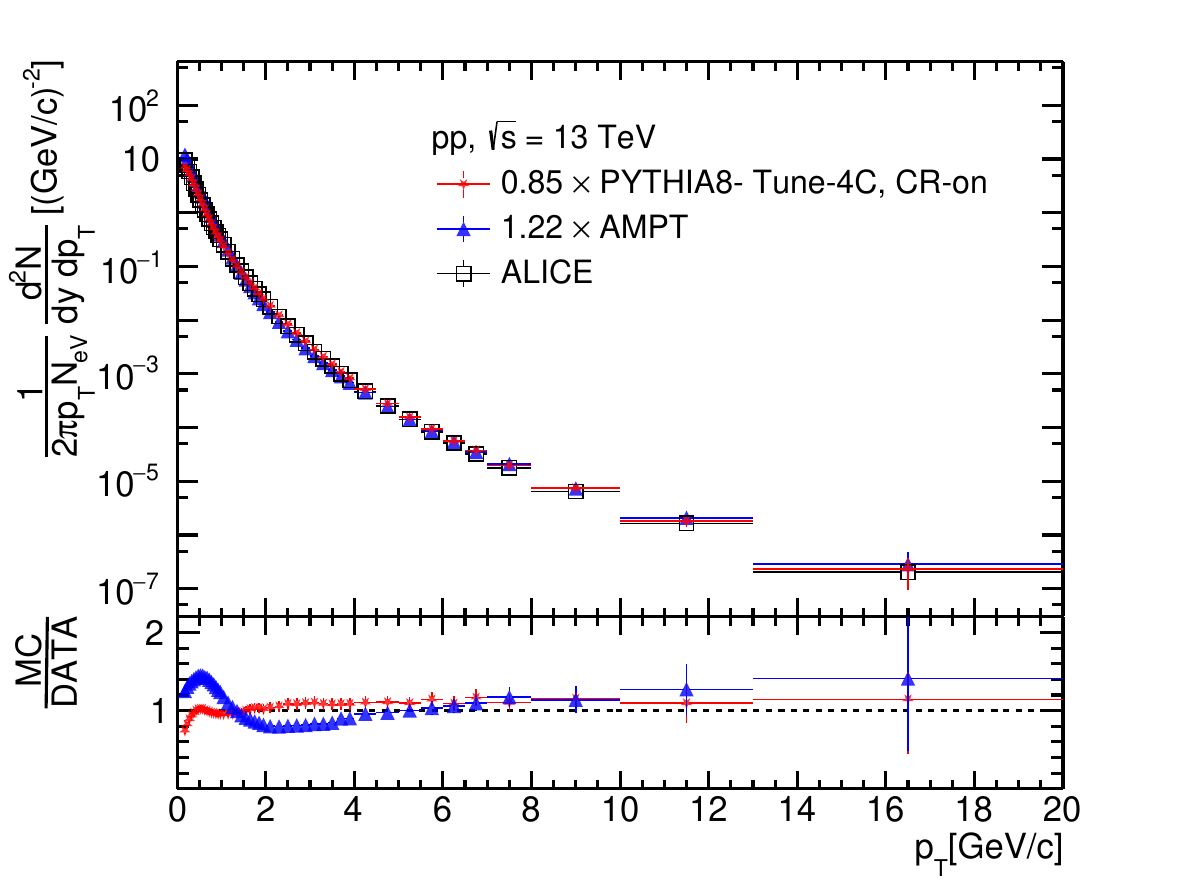}
\caption{(Color online) Comparison of PYTHIA8 generated data with AMPT and ALICE results \cite{ALICE:2015qqj} for transverse momentum spectra of all charged hadrons for minimum bias pp collisions at $\sqrt{s}$ = 13 TeV.}
\label{fig:datacomp}
\end{figure}

\begin{figure}[ht!]
\includegraphics[scale=0.4]{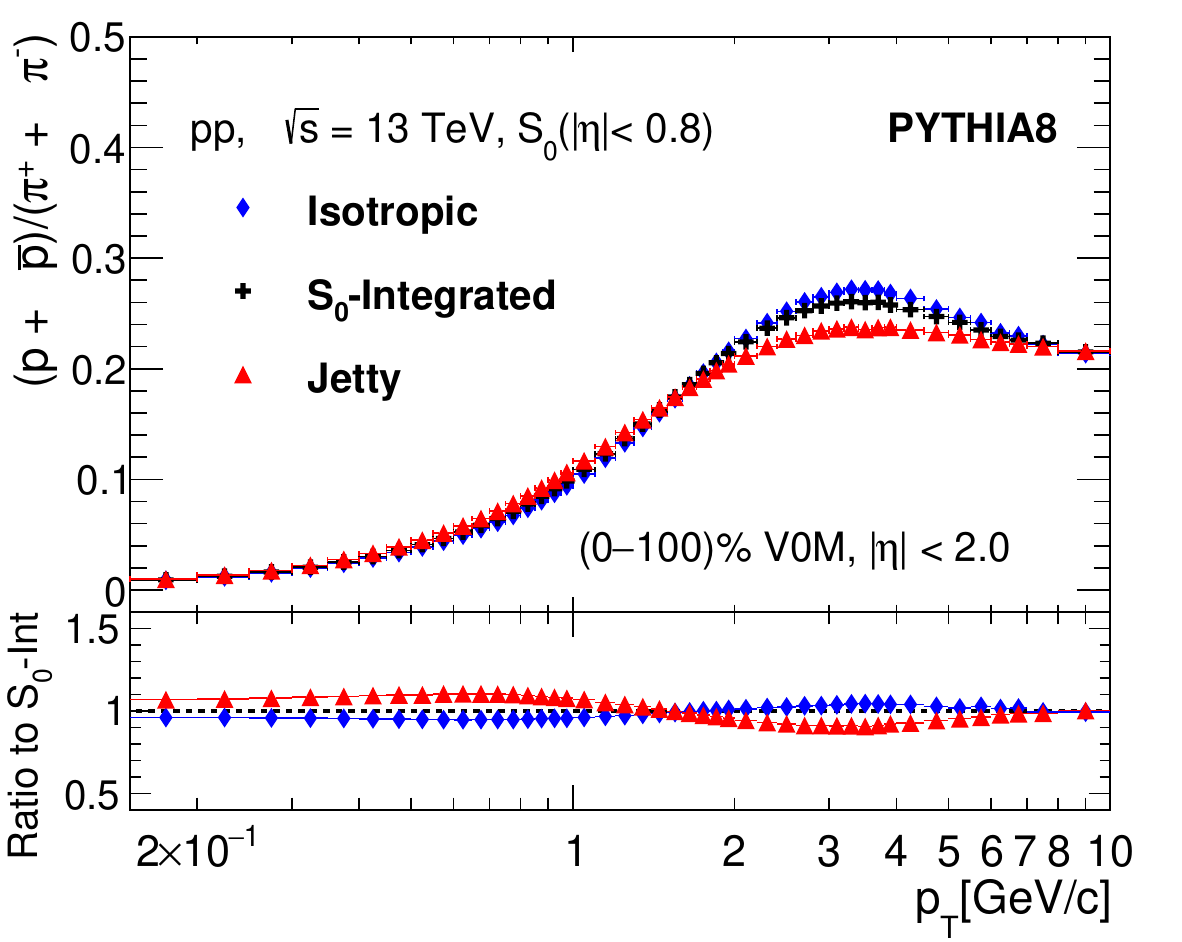}
\includegraphics[scale=0.4]{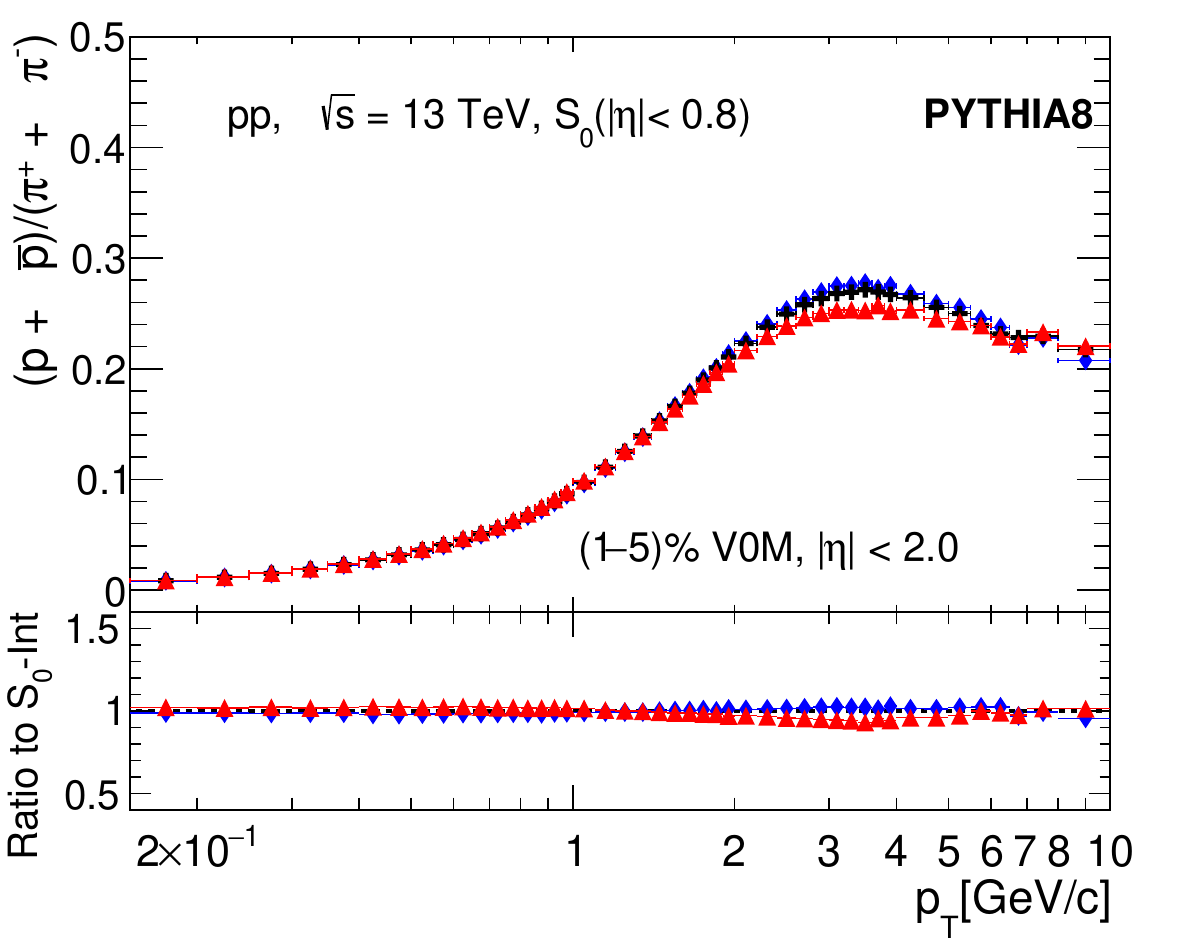}
\caption{(Color online) $p_{\rm T}$-differential proton to pion yield ratio as a function of $S_{0}(|\eta|<0.8)$ for (0-100)\% (upper) and (1-5)\% (lower) V0M classes in pp collisions at $\sqrt{s}=13$ TeV using PYTHIA8.}
\label{fig:particleratios_diffS0estimators}
\end{figure}

In Fig.~\ref{fig:particleratios_diffS0estimators}, we show $p_{\rm T}$-differential proton-to-pion yield ratio as a function of $S_{0}(|\eta|<0.8)$ for (0-100)\% (upper) and (1-5)\% (lower) V0M classes in pp collisions at $\sqrt{s}=13$ TeV using PYTHIA8. Here, one finds a similar qualitative behaviour of the proton-to-pion ratio as a function of $p_{\rm T}$ and transverse spherocity as shown in Fig.~\ref{fig:protonbypion}. However, in Fig.~\ref{fig:particleratios_diffS0estimators}, a weaker spherocity dependence is found in comparison to Fig.~\ref{fig:protonbypion} in both (0-100)\% and (1-5)\% V0M classes. This could result from a relatively weaker correlation of $S_{0}(|\eta|<0.8)$ with $\langle N_{\rm mpi}\rangle$, as shown in Fig.~\ref{fig:3S0comppThatNmpi}.

% \begin{figure*}[ht!]
% \includegraphics[scale=0.4]{RefereePlots/nMPI_0.8_S0_2.0_S0_4.0.pdf}
% \includegraphics[scale=0.4]{RefereePlots/nMPI_0.8_S0_2.0_S0_4.0_1percV0M.pdf}
% \caption{(Color online) }
% \label{fig:3S0compNmpi}
% \end{figure*}

\end{document}